\documentclass[10pt,twocolumn,letterpaper]{article}

\usepackage{iccv}
\usepackage{times}
\usepackage{epsfig}
\usepackage{graphicx}
\usepackage{amsmath}
\usepackage{amssymb}

\usepackage{xcolor}
\usepackage{soul}
\usepackage{makecell, subcaption, multirow}
\usepackage{booktabs}
\usepackage{gensymb}

\usepackage[pagebackref=true,breaklinks=true,letterpaper=true,colorlinks,bookmarks=false]{hyperref}

\iccvfinalcopy 

\def\hatx{\mathbf{\hat{x}}}
\def\hatc{\mathbf{\hat{c}}}
\def\x {\mathbf{x}}
\def\c {\mathbf{c}}
\def\y {\mathbf{y}}

\def\iccvPaperID{7434} 

\begin{document}

\title{Variational Deep Image Denoising}

\author{Jae Woong Soh\\
Department of ECE, INMC\\
Seoul National University\\
{\tt\small soh90815@ispl.snu.ac.kr}
\and
Nam Ik Cho\\
Department of ECE, INMC\\
Seoul National University\\
{\tt\small nicho@snu.ac.kr}
}

\maketitle

\begin{abstract}

Convolutional neural networks (CNNs) have shown outstanding performance on image denoising with the help of large-scale datasets. Earlier methods na\"{i}vely trained a single CNN with many pairs of clean-noisy images. However, the conditional distribution of the clean image given a noisy one is too complicated and diverse, so that a single CNN cannot well learn such distributions. Therefore, there have also been some methods that exploit additional noise level parameters or train a separate CNN for a specific noise level parameter. These methods separate the original problem into easier sub-problems and thus have shown improved performance than the na\"{i}vely trained CNN.
In this step, we raise two questions. The first one is whether it is an optimal approach to relate the conditional distribution only to noise level parameters. The second is what if we do not have noise level information, such as in a real-world scenario.
To answer the questions and provide a better solution, we propose a novel Bayesian framework based on the variational approximation of objective functions. This enables us to separate the complicated target distribution into simpler sub-distributions. Eventually, the denoising CNN can conquer noise from each sub-distribution, which is generally an easier problem than the original. Experiments show that the proposed method provides remarkable performance on additive white Gaussian noise (AWGN) and real-noise denoising while requiring fewer parameters than recent state-of-the-art denoisers.
\end{abstract}

\section{Introduction}
Image denoising has been an important task due to inevitable noise corruption in the image acquisition process.
It aims to reconstruct the underlying clean image from a noisy one. The observed noise from an imaging device is generally the accumulation of multiple noises from the sources such as capturing sensors and on-device image processing pipelines. Since the noise generation process is too complicated to be accurately modeled, the noise is usually assumed to be an AWGN based on the central limit theorem.

Recently, CNNs have shown great success in removing AWGN from images \cite{DnCNN, TNRD, RED, FFDNet, NLRN, N3}, largely surpassing traditional methods such as total variation, K-SVD denoising, NLM, BM3D, WNNM, etc. \cite{tv1, K-SVD, non-local1, BM3D, WNNM, TWSC}. However, earlier CNN-based methods were mostly non-blind ones, which need separate models trained (fitted) to different noise levels. When the input image bears a noise with a different level from the trained one, a severe performance drop occurs due to domain discrepancy between the distributions of training and test images. This phenomenon degrades its reliability and limits practical applications.

A na\"{i}ve approach to alleviating this problem is to compose training data consisting of images having a wide range of noise levels and train a single-model blind denoiser that can cope with broad noise levels \cite{DnCNN, MemNet, UNLNet}. However, the blind denoiser is generally not as good as the non-blind one that is well-fitted to the given noise level because learning a conditional mapping between the clean and noisy images with diverse distribution is more difficult than the learning of simpler distribution. Another approach is to develop flexible networks that can deal with multiple noise levels by exploiting additional information \cite{FFDNet, ATDNet}. In this case, a noise level estimator is additionally needed along with the denoising CNN, and hence we will refer to this approach as a two-stage system. One of its drawbacks is that they do not work properly when the noise level estimator fails to provide accurate information.

So far, it has been hypothesized that dividing a complex distribution into simpler sub-distributions will make a CNN easy to learn the overall task, eventually bringing performance gain. Also, providing additional information to a CNN is regarded as to ease the target distribution into simpler ones. For AWGN removal, the noise level has been provided as additional prior information without a doubt, which enables to separate the overall problem into sub-problems corresponding to specific noise levels. However, there can be a better way to divide the distribution, for example, by reflecting the image semantics \cite{SFT-GAN}.

Meanwhile, the distribution of noise from imaging devices referred to as ``real-noise,'' largely deviates from the i.i.d. Gaussian distribution. Hence, denoisers trained with Gaussian noise do not perform well for the real-noises, and thus they have limitations in practical, real-world situations.
Therefore, very recently, the real-noise denoising task attracted researchers, and several works have been proposed \cite{CBDNet, Unprocessing, RIDNet, VDN, CycleISP, AINDNet}.
Dealing with real-noise is a more challenging task than the AWGN in two aspects.

First, it is hard to build a paired clean-noisy dataset for real-noise denoising. Since the distribution of real-noise has more complicated features than the Gaussian, \emph{e.g.}, multi-modal, signal-dependent, spatially variant, etc., accurate modeling for real-world noises has also been a longstanding problem \cite{Noise1, Noise2, Noise3, Nam, Noise4, Physics}. Specifically, heteroscedastic or Poisson mixture models have been considered for reflecting some of these features \cite{Noise1, Noise2, Noise3, Noise4}. Recently, some researchers have taken many pairs of clean and real-noisy images with careful image acquisition settings \cite{SIDD, DND}. These methods can alleviate the data scarcity problem, but such acquisition processes are costly and labor-intensive. Some other works have modeled the noise corruption process based on the generative adversarial network (GAN) \cite{GCBD}, normalizing flows \cite{NoiseFlow}, or prior knowledge on camera pipeline and noise properties \cite{CBDNet, Unprocessing, Physics}.

Second, learning such a complex distribution may be a burdensome task to a single CNN. There have been some researches to address the previous issue, {\em i.e.,} precise real-noise modeling or acquiring well-registered real noisy-clean image pairs, but very few researchers considered how to relieve the hardness of training or make the best use of the dataset. Notably, most previous real-noise denoisers correspond to the category of na\"{i}ve blind denoisers.

To address the above issues, we propose a new method that can handle blind scenarios, including synthetic AWGN and real-world noise, namely Variational Deep Image Denoiser (VDID).
Our approach to solving the denoising problem is ``divide-and-conquer.'' We split the original objective into simpler sub-problems, which eventually ease the overall task.
For the Gaussian denoising, we raise doubt about conventional approaches that find different denoisers or control the features according to the noise level only. Instead, we seek an optimal data-driven prior or criterion to bring out the best performance with a single CNN.
For the real-noise denoising, we tackle the problem differently from the traditional methods.
Instead of looking for a good model for the real-noise distribution, we relax the original problem to simpler ones by dividing the underlying complex posterior distribution into sub-modal distributions.
Specifically, we formulate our objective in terms of maximum a posterior (MAP) inference and present an approximated form of the objective by introducing a latent variable based on variational Bayes. By doing so, the network learns its latent space, which represents the sub-distributions of the noisy images. In other words, we divide the original problem into several sub-problems and solve each case separately. In particular, we introduce a latent representation of noisy images and exploit their representations as an additional prior to handle them differently. Furthermore, our method is trained in an end-to-end scheme without any additional noise information.

In summary, our contributions are as follows.
\begin{itemize}
\setlength\itemsep{-0.1em}
	\item We present a novel CNN-based blind image denoiser, which is trained in an end-to-end scheme.
	\item To the best of our knowledge, we first tackle the image denoising as to relaxing the original problem into easier ones.
	\item Based on the variational approximation, we reformulate our target problem to include the auto-encoding term, which incorporates underlying noisy image distribution.
	\item Based on the latent space implying the noisy image manifolds, our VDID can focus on simpler sub-distributions of the original problem.
	\item From the extensive experiments, we have shown that the proposed method achieves state-of-the-art performances while requiring fewer parameters.
\end{itemize}

\section{Related Work}

Based on the assumption of image noise distribution as pixel-wise i.i.d. Gaussian with a standard deviation of $\sigma_N$, AWGN denoising has been a longstanding problem. Recently, CNN-based denoisers have also been actively studied \cite{DnCNN, TNRD, RED, FFDNet, ATDNet, NLRN, N3}, and noise removal in real-world images attracted researchers due to its practical importance \cite{GCBD, CBDNet, Unprocessing, NoiseFlow, CycleISP, Physics, DND, SIDD, RIDNet, AINDNet}.
From the viewpoint of objectives, we divide former related approaches into three categories: Specific Non-blind Model, Na\"{i}ve Blind Model, and Two-Stage Blind Model. For the entire paper, we will denote the noisy image as $\y$, the clean image $\x$, and the underlying data distribution $p_{data}$.

\subsection{Specific Non-blind Model}

Most CNN-based AWGN denoisers such as DnCNN-S, RED, and NLRN \cite{DnCNN, RED, NLRN} adopt ``specific non-blind model,'' which is a separate network trained for a specific noise level. Its objective can be expressed as
\begin{equation}
\hat{\theta_i}=\arg \max_{\theta_i} \mathbb{E}_{p_{data}(\x, \y, \sigma_N)}[\log p_{\theta_i}(\x|\y,\sigma_N^i)],
\end{equation}
where $\theta_i$ denotes the parameters of the $i$-th network. Specifically, most previous works have taken one of two choices for the noise levels: $\sigma_N^i \in \{15,25,50\}$ or $\{10,30,50,70\}$. At the inference, the noise level information of a test image is required for obtaining a desirable output. Also, a number of networks should be prepared in the bag, requiring a large memory space.

\subsection{Na\"{i}ve Blind Model}
Most blind denoisers are in this category of na\"{i}ve blind model, totally relying on the representation power of CNN. Especially, since exploiting the information such as noise level is very difficult in the case of real-noises, it would be an appropriate approach to adopt the na\"{i}ve blind model. The objective of this approach is expressed as
\begin{equation}
\hat{\theta}=\arg \max_{\theta} \mathbb{E}_{p_{data}(\x, \y)}[\log p_{\theta}(\x|\y)],
\end{equation}
for training a single network parameterized by $\theta$ to capture the conditional distribution. In general, na\"{i}ve blind models show worse performance than the specific models \cite{DnCNN, OneSize}, since the distribution $p(\x|\y)$ is more complicated than $p(\x|\y,\sigma_N)$. Specifically, DnCNN-B, UNLNet, GCBD, and RIDNet \cite{DnCNN, UNLNet, GCBD, RIDNet} come into this category.

\subsection{Two-Stage Blind Model}
Some recent methods adopt the two-stage blind model, where noise level parameters $\c$ are first estimated and then fed to the denoising network. Its objective is
\begin{align}
\hat{\phi} =& \arg \max_{\phi} \mathbb{E}_{p_{data}(\y, \c)}[ \log p_{\phi}(\c|\y)],\\
\hat{\theta} =&\arg \max_{\theta} \mathbb{E}_{p_{data}(\x, \y, \c)} [\log p_{\theta}(\x|\y,\c)]],
\end{align}
where $\phi$ and $\theta$ denote the parameters of noise estimator and denoiser, respectively.
For the Gaussian noise, $\c$ is selected as the standard deviation of Gaussian distribution \cite{FFDNet, ATDNet}, and for the real-noise, more complicated parameters are selected \cite{CBDNet, VDN, AINDNet}. Notably, prior information on noise is additionally required for this setting.

\section{Variational Deep Image Denoising}

This section presents our objective in terms of MAP inference and then reformulates the problem to tractable sub-problems.
\subsection{Problem Statement}
Given a noisy image $\mathbf{y}$, the objective is to find a latent clean image $\mathbf{x}$. The MAP inference for this problem is
\begin{equation}
\hatx = \arg \max_\x \log p(\x|\y).
\end{equation}
Most of the traditional approaches divide the posterior term into likelihood and prior terms as
\begin{align}\nonumber
\hspace{-0.3cm}\arg \max_\x \log p(\x|\y) &= \arg \max_\x \log p(\y|\x) + \log p(\x)\\
&= \arg \min_\x \frac{1}{2\sigma^2} ||\y-\x||_2^2 + \Phi(\x),
\end{align}
and solve it with a well-designed prior $\Phi(\x)$ under the i.i.d. Gaussian assumption. However, we do not follow this approach since we use a data-driven discriminative learning scheme. Also, it needs to be noted that the traditional approaches require a complex data likelihood term, which is hard to approximate the likelihood of real noisy images.

We first introduce a new latent random variable $\c$, which implies suitable information both for denoising (task-relevant information) and properties of clean/noisy image (domain-relevant information).
Then, we bring an inference problem of the posterior $p(\c|\x, \y)$, in which the latent $\c$ includes both the domain- and task-relevant information learned from clean and noisy images. However, this inference problem is intractable. Also, our other objective is to infer $\x$, which cannot be observed during the inference.
In summary, our interested inference problems are $\log p(\c|\x,\y)$ and $\log p(\x|\y)$, which are intractable or unobservable.

\subsection{Proposed Variational Lower Bound}
To approximate the posterior $p(\c|\x, \y)$, we introduce a tractable probability distribution $q(\c|\y)$.
Then, the joint probability distribution $\log p(\x,\y)$ can be reformulated as
\begin{align}\nonumber
\log p(\x, \y) =& \mathbb{E}_{\c \sim q(\c|\y)} [\log p(\x|\y,\c)]\\\nonumber
& + D_{KL}(q(\c|\y)||p(\c|\x,\y))\\\nonumber
& - D_{KL}(q(\c|\y)||p(\c))\\
& + \mathbb{E}_{\c \sim q(\c|\y)} [\log p(\y|\c)],
\end{align}
with some prior distribution $p(\c)$.
To approximate the intractable KL divergence term between $q(\c|\y)$ and $p(\c|\x, \y)$, we introduce a \emph{variational lower bound} $\mathcal{L}$.

\noindent\textbf{Definition 1}
\emph{Variational lower bound} $\mathcal{L}$ is defined as
\begin{align}\nonumber
\mathcal{L} =& \mathbb{E}_{\c \sim q(\c|\y)} [\log p(\x|\y,\c)]\\
& - D_{KL}(q(\c|\y)||p(\c)) + \mathbb{E}_{\c \sim q(\c|\y)} [\log p(\y|\c)].
\end{align}

\noindent\textbf{Theorem 1}
\textit{Given a noisy image $\y$ and its underlying clean image $\x$, the joint log-distribution $\log p(\x,\y)$ can be reformulated including variational lower bound $\mathcal{L}$ as}
\begin{equation}
\log p(\x,\y) = \mathcal{L} + D_{KL}(q(\c|\y)||p(\c|\x,\y)).
\end{equation}
Then,
\begin{equation}
\log p(\x,\y) \geq \mathcal{L}.
\end{equation}
(The overall derivation and proof can be found in \emph{appendix}.)

\noindent\textbf{Definition 2}
We define a log-posterior $q(\x|\y)$ which approximates the original posterior $p(\x|\y)$ as
\begin{align}\nonumber
q(\x|\y) &= \int_\c q(\x,\c|\y) d\c\\
&= \int_\c p(\x|\y,\c)q(\c|\y) d\c.
\end{align}
Then, the MAP inference given $\y$ can be done by $\arg \max_\x \log q(\x|\y)$.
\\

Through our reformulation, maximizing the joint probability, and the objective to minimize the KL divergence between our variational distribution and the posterior on $\c$, are simultaneously approximated as to maximize our \emph{variational lower bound}.
Notably, the first term of $\mathcal{L}$ is the only term relevant to the relation between $\x$ and $\y$, responsible for denoising or reconstruction.
The other terms are regularization terms, which impose constraints on the latent variable $\c$. The second term is the KL divergence, which constrains the latent distribution, and the third term is the auto-encoder reconstruction term of noisy images.

As neural networks are experts in inference, in an amortized way \cite{amortized, AVI}, we employ three CNNs parameterized by $\theta$, $\phi_E$, and $\phi_D$ for the variational inference. Specifically, our final objective is
\begin{align}
\arg \max_{\theta, \phi_E, \phi_D} &\mathbb{E}_{p_{data}(\x, \y)}[\nonumber
\mathbb{E}_{\c \sim q_{\phi_E}(\c|\y)} [\log p_{\theta}(\x|\y,\c)]\\\nonumber
& - D_{KL}(q_{\phi_E}(\c|\y)||p(\c)) \\
& + \mathbb{E}_{\c \sim q_{\phi_E}(\c|\y)} [\log p_{\phi_D}(\y|\c)]],
\end{align}
with the underlying empirical data distribution $p_{data}(\x, \y)$.
By introducing such regularizations, our denoiser can approximately solve a MAP problem according to the latent $\c$, where $\c$ is the variable involved in the noisy image generation process. In other words, our denoiser divides the problem according to the latent $\c$, where $\c$ should ``imply'' the noisy image manifold.

\subsection{Network Architecture}

\begin{figure}
	\centering
	\includegraphics[width=1.0\linewidth]{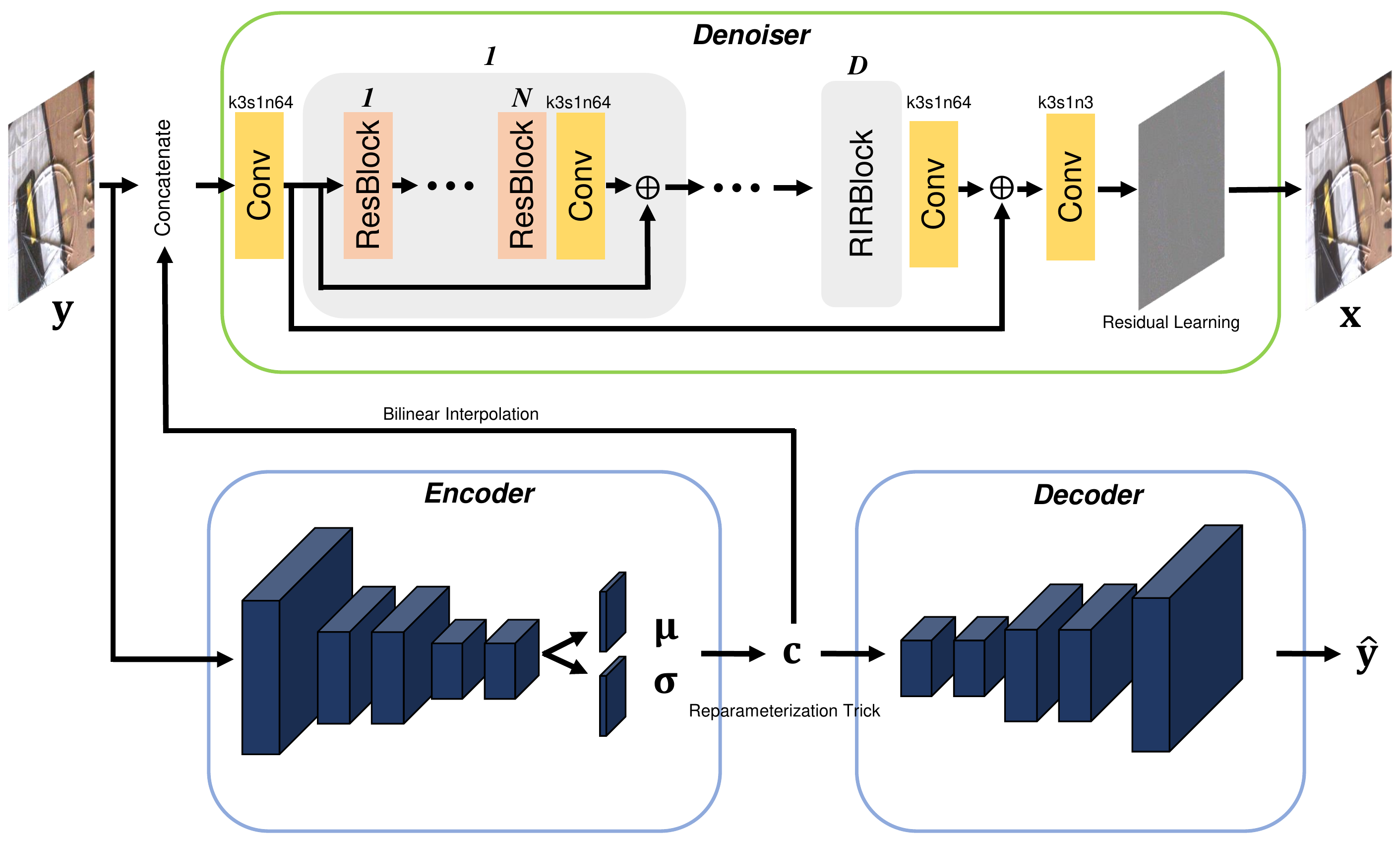}
	\vspace{-0.3cm}
	\caption{The overall architecture of proposed VDID, where $k$, $s$, and $n$ denote kernel size, stride, and the number of filters, respectively. 
	}
	\label{fig:network}
\end{figure}

The overall network architecture is shown in \figurename{~\ref{fig:network}}. The denoiser takes a noisy image $\y$ with the latent variable $\c$ concatenated along the channel axis to infer the clean image $\x$. The denoiser is fully convolutional, thus highly scalable. For the denoiser, the residual block (ResBlock) is adopted as the basic building block \cite{ResNet, EDSR}. Precisely, the same residual block of \cite{EDSR} is adopted, which consists of $3\times 3$ convolution layers of $64$ filters followed by the rectified linear unit (ReLU) and another convolution layer (Conv-ReLU-Conv). Then, the input is added to the output of the convolution layer, which forms the skip-connection. The $N$ number of residual blocks and one convolution layer compose the residual-in-residual block (RIRBlock) \cite{RCAN}. The denoiser consists of $D$ RIRBlocks with some convolution layers and a long skip-connection, as shown in \figurename{~\ref{fig:network}}. The last convolution layer infers the residual image (noise) instead of the clean image itself, according to \cite{DnCNN}.

The encoder and decoder networks are simple feedforward convolutional networks without skip-connection. The encoder decreases the feature map's spatial size twice (one-fourth of its height and width), and the output $\c$ has four channels. For the differentiable Monte Carlo, we adopt the reparameterization trick \cite{VAE, BetaVAE} as
\begin{equation}
\epsilon \sim \mathcal{N}(0, I),\quad \c = \epsilon \odot \sigma + \mu,
\end{equation}
where 
$\odot$ denotes Hadamard product, and $\mu$ and $\sigma$ are the encoder's outputs. The decoder network has symmetrical architecture as the encoder. The details of the architectures are presented in \emph{appendix}.

\subsection{Corresponding Loss Terms}
\paragraph{First Term}
For the first term of our objective, which is the denoising term, we adopt mean absolute error (MAE) between the ground-truth clean image and the inferred output \cite{EDSR} to minimize the distortion. We denote the corresponding loss term as $\mathcal{L}_{denoise}$:
\begin{equation}
\mathcal{L}_{denoise}= ||\x-\hat{\x}||_1,
\end{equation}
where $\hat{\x}$ denotes the output of the denoiser.

\vspace{-0.3cm}
\paragraph{Second Term}
The KL divergence between the prior distribution and the posterior can be calculated analytically. We set the prior $p(\c)$ as Gaussian distribution with zero mean and unit covariance. Thus, the KL divergence term is
\begin{equation}
D_{KL}(q(\c|\y)||p(\c))= D_{KL}(\mathcal{N}(\mu, \Sigma)||\mathcal{N}(0, I)).
\end{equation}


\vspace{-0.3cm}
\paragraph{Third Term}
As the third term, which is the auto-encoder reconstruction term, we first minimize MAE between the noisy input image and the decoder's output. However, using only pixel-wise loss strictly assumes $p(\y|\c)$ to be a family of the probability distribution of Laplacian or Gaussian. To relax and better learn the noisy image distribution, we adopt additional adversarial loss \cite{GAN}. Specifically, we adopt non-saturating GAN loss \cite{GAN}, corresponding to minimizing Jensen-Shannon divergence between $p_{data}(\y)$ and $p_{\phi_{D}}(\y|\c)$. For the AWGN removal, we add our known prior that the latent space should include the noise level information. Hence, we add a noise level estimation loss.
Eventually, the latent $\c$ works as a prior for the denoiser, which contains an abstract of noisy image distribution.

The corresponding loss term $\mathcal{L}_{recon}$ is described as
\begin{equation}
\mathcal{L}_{recon}= ||\y-\hat{\y}||_1 + \lambda_1 \mathcal{L}_{adv} + \lambda_2 ||\sigma_N - EST(\c)||_1
\end{equation}
where $\hat{\y}$ denotes the output of the decoder, and $EST(\cdot)$ is a simple two-layer CNN (conv-relu-conv).

\vspace{-0.3cm}
\paragraph{Overall Loss}
The overall loss is the sum of three terms, with the normalized KL divergence multiplied by $\beta$ \cite{BetaVAE},
\begin{equation}
\arg \min_{\theta, \phi_E, \phi_D}  \mathbb{E}_{p_{data}(\x, \y)}[\mathcal{L}_{denoise} + \beta  D_{KL} + \mathcal{L}_{recon}].
\end{equation}

\subsection{Discussions}

\paragraph{Probabilistic View}
Let us assume the denoising loss term as L2-norm for simplicity, which assumes i.i.d. Gaussian distribution as target probabilistic family (L1-norm is associated with Laplacian distribution). We refer a neural network parameterized by $\theta$ as $f_{\theta}(\cdot)$.

The na\"{i}ve blind model models $p_{\theta}(\x|\y) = \mathcal{N}(f_{\theta}(\y),I/2)$, and learning the data distribution under this family hinders expressiveness. Since the posterior distribution $p(\x|\y)$ including diverse degradation is too complex to be captured by a single Gaussian, its performance cannot be expected as much as the specific model with simpler $p(\x|\y)$ with a single degradation.
On the other hand, our framework models $p(\x|\y, \c) = \mathcal{N}(f_{\theta}(\y, \c), I/2)$, and it is still the Gaussian form. But it learns different mean values with respect to $\c$, which grants more representation power by learning multiple Gaussians in accordance with $\c$.
Then, the marginal posterior has more representative power $p(\x|\y) = \int_{\c} p(\x|\y,\c),p(\c|\y) d\c$.
Though, our inference approximates the marginal posterior through Monte-Carlo using only one sample of $\c$.

The two-stage model can be considered a special case of our method where $q(\c|\y)$ is chosen as deterministic.
In this case, the $c$ is determined based on the prior knowledge and carefully modeled by ``understanding the data.''
Then, the point estimate of $\hatc = \arg \max_\c q(\c|\y)$ is used for the second step inference \cite{DUBD}.
Note that this bi-level optimization scheme would be sub-optimal to the task objective compared to the joint optimization.
Unlike the two-stage model, our method is more Bayesian and implicitly learns $\c$, which is enforced to contain the degradation information along with the original image content information. In other words, our method conducts Bayesian inference whereas the two-stage model conducts deterministic estimates, and also the additional information $\c$ is learned by enforcing the network to ``understand the data.''

\vspace{-0.3cm}
\paragraph{Connection to Blind AWGN Denoiser}
For the AWGN, where $q(\c|\y)$ is chosen as deterministic, the proposed method can be regarded as a two-stage blind denoiser, \emph{i.e.}, a noise level estimator (corresponding to the encoder) and a flexible denoiser (corresponding to the denoiser) that works for a range of noise levels. Hence, a two-stage blind denoiser can be considered a special case of our approach.
In this case, the $\c$ is carefully determined by ``understanding the data.'' For AWGN, it is chosen as a standard deviation of Gaussian distribution, based on the prior knowledge.
The point estimate of $\hatc = \arg \max_\c q(\c|\y)$ is used for the second step inference \cite{DUBD}.
Note that this bi-level optimization scheme would be sub-optimal to the task objective, compared to the joint optimization.
On the other hand, our method implicitly learns $\c$, which enforces the network to ``understand the data.''

\vspace{-0.3cm}
\paragraph{Discussions on Loss Terms}
When we use only the first term, it is just a na\"{i}ve approach to solve a blind denoising problem, totally relying on the discriminative power of CNN. In this case, the objective is
\begin{equation}
\arg \max_{\theta, \phi_E} \mathbb{E}_{p_{data}(\x, \y)}[\mathbb{E}_{\c \sim q_{\phi_E}(\c|\y)} [\log p_{\theta}(\x|\y,\c)]].
\end{equation}
It is notable that no matter what latent distribution $q_{\phi_E}(\c|\y)$ we choose, this criteria is maximized if for each $\c$, $\mathbb{E}_{p_{data}(\x,\y)}[\log p_\theta(\x|\y,\c)]$ is maximized. In other words, there is a trivial solution ``independent'' to $\c$, if our model has the optimal parameter satisfying $\theta^*=\arg \max_\theta \mathbb{E}_{p_{data}(\x, \y)}[\log p_\theta(\x|\y)]$. In this case, $\c$ collapses and the original problem cannot be divided into sub-problems. The second term gives regularization constraints, where the KL divergence term forces disentanglement of $\c$, giving ``discriminative power'' over observed noisy images $\y$.
The third term further gives constraints on the latent variable $\c$. As the auto-encoder reconstruction term forces the reconstruction from $\c$ to $\y$, this term forces $\c$ to include the information on a noisy image.

\begin{table*}[!t]
	\caption{The average PSNR on AWGN denoising. The best results are highlighted in \textcolor{red}{red} and the second best in \textcolor{blue}{blue}.}
	\vspace{-0.5cm}
	\begin{center}
		\resizebox{0.9\linewidth}{!}{
			\begin{tabular}{c|c|c|c|c|c|c|c|c}
				\Xhline{4\arrayrulewidth}
				\rule[-1ex]{0pt}{3.5ex}
				Noise level & Dataset & CBM3D \cite{BM3D} & RED \cite{RED} & CDnCNN \cite{DnCNN} & FFDNet \cite{FFDNet} & UNLNet \cite{UNLNet} & VDN \cite{VDN} & VDID (Ours)\\
				\hline\hline
				
				\rule[-1ex]{0pt}{3.5ex}
				\multirow{3}{*}{$\sigma_N = 10$}&CBSD68&35.91&33.89&36.13&36.14& 36.20 & \textcolor{blue}{36.29} & \textcolor{red}{36.34}\\
				\rule[-1ex]{0pt}{3.5ex}
				&Kodak24  & 36.43 & 34.73 & 36.46 & 36.69 & - & \textcolor{blue}{36.85} & \textcolor{red}{37.02}\\
				\rule[-1ex]{0pt}{3.5ex}
				&Urban100& 36.00 & 34.42 & 34.61 & 35.78 & - & \textcolor{blue}{35.97} & \textcolor{red}{36.30} \\
				\hline\hline
				
				\rule[-1ex]{0pt}{3.5ex}
				\multirow{3}{*}{$\sigma_N = 30$}& CBSD68 & 29.73 & 28.45 & 30.34 & 30.32 & 30.21 & \textcolor{red}{30.64} & \textcolor{red}{30.64} \\
				\rule[-1ex]{0pt}{3.5ex}
				&Kodak24  & 30.75 & 29.53 & 31.17 & 31.27 & 31.18 & \textcolor{blue}{31.67} & \textcolor{red}{31.74} \\
				\rule[-1ex]{0pt}{3.5ex}
				&Urban100& 30.36& 28.84 & 30.00 & 30.53 & 30.41 & \textcolor{blue}{31.14} & \textcolor{red}{31.41}\\
				\hline\hline
				
				\rule[-1ex]{0pt}{3.5ex}
				\multirow{3}{*}{$\sigma_N = 50$}&CBSD68& 27.38& 26.34 & 27.95 & 27.97 & 27.85 & \textcolor{red}{28.33} & \textcolor{red}{28.33} \\
				\rule[-1ex]{0pt}{3.5ex}
				&Kodak24  & 28.46 & 27.42  & 28.83 & 28.98 & 28.86 & \textcolor{blue}{29.44} & \textcolor{red}{29.49} \\
				\rule[-1ex]{0pt}{3.5ex}
				&Urban100& 27.94 & 26.25 & 27.59 & 28.05 & 27.95 & \textcolor{blue}{28.86} &  \textcolor{red}{29.10}\\
				\hline\hline
				
				\rule[-1ex]{0pt}{3.5ex}
				\multirow{3}{*}{$\sigma_N = 70$}&CBSD68& 26.00 & 25.09 & 25.66 & 26.55 & - & \textcolor{blue}{26.93} & \textcolor{red}{26.94}\\
				\rule[-1ex]{0pt}{3.5ex}
				&Kodak24  & 27.09 & 26.16 & 26.36 & 27.56 & - & \textcolor{blue}{28.05} & \textcolor{red}{28.10} \\
				\rule[-1ex]{0pt}{3.5ex}
				&Urban100& 26.31 & 24.58 &  25.24 & 26.40 & - & \textcolor{blue}{27.31} & \textcolor{red}{27.55} \\
				
				\Xhline{4\arrayrulewidth}
			\end{tabular}
		}
	\end{center}	
	\label{tab:AWGN}
\end{table*}

\section{Experimental Results}
We perform denoising experiments on Gaussian and real noises. For our VDID, we set $N=5$ and $D=5$, which amounts to about $2.2$ M parameters, including the denoiser and the encoder. All the results are evaluated in sRGB space and demonstrated with PSNR and SSIM \cite{SSIM}.

\subsection{Implementation Details}
For the training, we extract patches with the size of $96 \times 96$ from training images for AWGN and $256 \times 256$ for real-noise. We adopt Adam optimizer with $\beta_1 = 0.9$ and $\beta_2=0.999$. For the loss term, we set $\beta=0.01, \lambda_1=0.001$, and $\lambda_2=1$ for AWGN, $\lambda_2=0$ for real-noise. For data augmentation, a random flip and $90 \degree$ rotations of the patches are applied. The initial learning rate is $2\times 10^{-4}$ and decayed half in every $100,000$ iterations, until it reaches $2\times 10^{-5}$. The batch size is set to $32$ for AWGN and $4$ for real-noise.

\subsection{Results on AWGN Removal}
For training, we use DIV2K \cite{DIV2K} training dataset which includes $800$ high-resolution images, and add synthetic Gaussian noise with noise level $\sigma_N \in [5, 70]$. The performance is evaluated with three color-image datasets: CBSD68 \cite{BSD}, Kodak24, and Urban100 \cite{Urban} with noise levels $\sigma_N = 10,~30,~50,~70$. We compare our method with several AWGN denoising algorithms: CBM3D \cite{BM3D}, RED \cite{RED}, CDnCNN \cite{DnCNN}, FFDNet \cite{FFDNet}, UNLNet \cite{UNLNet}, and VDN \cite{VDN}. The results are presented in \tablename{~\ref{tab:AWGN}}.

Note that CBM3D \cite{BM3D}, RED \cite{RED}, and FFDNet \cite{FFDNet} are non-blind methods, whereas the rest are blind ones.
In most cases, VDN \cite{VDN} and our VDID show the best PSNR results, but we note that our VDID needs a smaller number of parameters (2.2 M) compared to VDN \cite{VDN} (7.8 M). In conclusion, the results show that our VDID surpasses other methods considering the tradeoff between the performance and the number of parameters.

\subsection{Results on Real-Noise Removal}
We use training images of Smartphone Image Denoising Dataset (SIDD) \cite{SIDD}, which is a collection of pairs of noisy and clean images from five smartphone cameras. It consists of $320$ image pairs for training. To augment more datasets of synthesized images, we also used DIV2K \cite{DIV2K} training images, which includes $800$ high-resolution images. To generate noisy images, we adopt the noise synthesis process of CBDNet \cite{CBDNet}. We compare with several image denoising methods: BM3D \cite{BM3D}, WNNM \cite{WNNM}, DnCNN \cite{DnCNN}, TNRD \cite{TNRD}, FFDNet \cite{FFDNet}, GCBD \cite{GCBD}, CBDNet \cite{CBDNet}, RIDNet \cite{RIDNet}, VDN \cite{VDN}, and AINDNet \cite{AINDNet}.
For the evaluation, we use two widely-used real image denoising benchmarks.

\begin{itemize}
	\item \textbf{SIDD:}
	SIDD provides $1,280$ small patches for validation and $1,280$ for test benchmark, which are visually similar to training images. Ground-truth patches for the validation set are provided, but not for the test set.
	\item \textbf{DND:}
	Darmstadt Noise Dataset (DND) consists of $50$ images with real-noise from $50$ scenes from four consumer cameras. Then, the images are further cropped by the provider, which results in $1,000$ small patches with a size of $512 \times 512$.
	
\end{itemize}

\begin{table}
	\caption{Results on SIDD \cite{SIDD} benchmark. The best results are highlighted in \textbf{bold}.}
	\label{tab:SIDD}
	\vspace{-0.3cm}
	\centering
	\resizebox{1.0\linewidth}{!}{
		\begin{tabular}{lcccc}
			\Xhline{4\arrayrulewidth}
			\rule[-1ex]{0pt}{3.5ex}
			Method & Blind/Non-blind & Parameters & PSNR & SSIM\\
			\hline\hline
			\rule[-1ex]{0pt}{3.5ex}
			BM3D \cite{BM3D} & Non-blind & - & 25.65 & 0.685 \\
			\rule[-1ex]{0pt}{3.5ex}
			WNNM \cite{WNNM} & Non-blind & - & 25.78 & 0.809 \\
			\rule[-1ex]{0pt}{3.5ex}
			DnCNN \cite{DnCNN} & Non-blind & 668 K & 23.66 & 0.583 \\
			\rule[-1ex]{0pt}{3.5ex}
			TNRD \cite{TNRD} & Non-blind & 27 K & 24.73 & 0.643 \\
			\rule[-1ex]{0pt}{3.5ex}
			CBDNet \cite{CBDNet} & Blind & 4.4 M & 33.28 & 0.868 \\
			\rule[-1ex]{0pt}{3.5ex}
			RIDNet \cite{RIDNet} & Blind & 1.5 M & 38.71 & 0.914 \\
			\rule[-1ex]{0pt}{3.5ex}
			VDN \cite{VDN} & Blind & 7.8 M & 39.26 & 0.955 \\
			\rule[-1ex]{0pt}{3.5ex}
			AINDNet+TF \cite{AINDNet} & Blind & 13.7 M & 38.95 & 0.952 \\
			\hline
			\rule[-1ex]{0pt}{3.5ex}
			VDID (Ours)& Blind & 2.2 M & 39.25 & 0.955 \\
			\rule[-1ex]{0pt}{3.5ex}
			VDID+ (Ours)& Blind & 2.2 M & \textbf{39.33} & \textbf{0.956}\\
			\Xhline{4\arrayrulewidth}
		\end{tabular}
	}
\end{table}

\begin{table}
	\caption{Results on DND \cite{DND} benchmark. The best results are highlighted in \textbf{bold}.}
	\label{tab:DND}
	\vspace{-0.3cm}
	\centering
	\resizebox{1.0\linewidth}{!}{
		\begin{tabular}{l|cccc}
			\Xhline{4\arrayrulewidth}
			\rule[-1ex]{0pt}{3.5ex}		
			Method & Blind/Non-blind & Parameters & PSNR & SSIM \\
			\hline\hline
			\rule[-1ex]{0pt}{3.5ex}
			BM3D \cite{BM3D} & Non-blind & - & 34.51 & 0.8507 \\ 
			\rule[-1ex]{0pt}{3.5ex}
			WNNM \cite{WNNM} & Non-blind & - & 34.67 & 0.8646 \\
			\rule[-1ex]{0pt}{3.5ex}
			DnCNN+ \cite{DnCNN} & Non-blind & 668 K & 37.90 & 0.9430 \\
			\rule[-1ex]{0pt}{3.5ex}
			FFDNet+ \cite{FFDNet} & Non-blind & 825 K & 37.61 & 0.9415 \\
			\rule[-1ex]{0pt}{3.5ex}
			GCBD \cite{GCBD} & Blind & 561 K & 35.58 & 0.9217 \\
			\rule[-1ex]{0pt}{3.5ex}
			CBDNet \cite{CBDNet} & Blind & 4.4 M & 38.06 & 0.9421 \\
			\rule[-1ex]{0pt}{3.5ex}
			RIDNet \cite{RIDNet} & Blind & 1.5 M & 39.26 & 0.9528 \\
			\rule[-1ex]{0pt}{3.5ex}
			VDN \cite{VDN} & Blind & 7.8 M & 39.38 & 0.9518 \\
			\rule[-1ex]{0pt}{3.5ex}
			AINDNet(S) \cite{AINDNet} & Blind & 13.7 M & 39.53 &\textbf{0.9561} \\
			\hline
			\rule[-1ex]{0pt}{3.5ex}
			VDID (Ours) &Blind & 2.2 M & 39.63 & 0.9528 \\
			\rule[-1ex]{0pt}{3.5ex}
			VDID+ (Ours)&Blind & 2.2 M & \textbf{39.69} & 0.9532\\
			
			\Xhline{4\arrayrulewidth}			
		\end{tabular}
	}
\end{table}

Overall quantitative comparisons on two benchmarks are listed in \tablename{~\ref{tab:SIDD}} and \ref{tab:DND}. It is observed that our method shows better results than others, on both benchmarks in terms of PSNR and SSIM \cite{SSIM}. Note that we also demonstrate results with self-ensemble \cite{Sevenways} based on geometric transformation, which is denoted with `+' sign.


For a fair comparison, we also denote the number of parameters for the CNN-based methods.
Since our VDID solves simpler sub-problems conditioned on the latent variable $\c$, it requires a smaller network than other denoisers based on na\"{i}ve blind setting. In other words, the problem given to other methods is more complicated due to the difficulty of ill-posed real image denoising. Thus, our method achieves state-of-the-art performances in real-world image denoising while requiring fewer parameters than others.

\subsection{Visualized Results}

For qualitative evaluation, we present a visual comparison in \figurename{~\ref{fig:001}. It shows that denoisers trained with AWGN, such as DnCNN \cite{DnCNN} and FFDNet \cite{FFDNet}, fail to remove the noise or tend to over-smooth the result, suffering from the discrepancy between the target noise distributions. On the other hand, methods for real-noise removal show more plausible results. However, most methods tend to over-smooth the patterns as shown in the green box, whereas our VDID shows better-restored line patterns. Also, our method shows clearer results in the text region of the red box.
Through the overall visualization, our method shows visually pleasing results on real-world images. More visualized comparisons are presented in \emph{appendix}.

\begin{figure}	
	\begin{center}		
		\captionsetup{justification=centering}
		\begin{subfigure}[t]{0.24\linewidth}
			\centering
			\includegraphics[width=1\columnwidth]{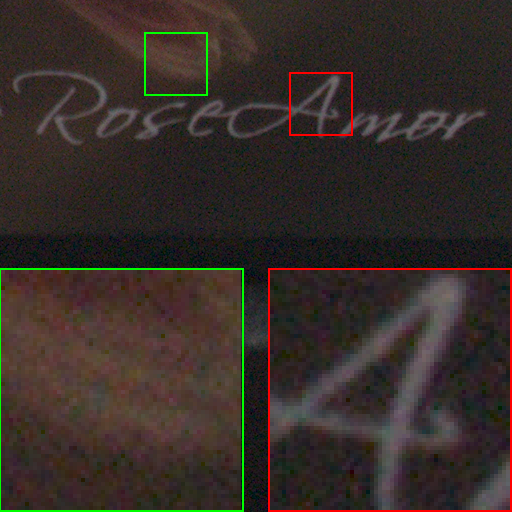}
			\caption*{Noisy \\ 30.42/0.5997}
		\end{subfigure}
		\begin{subfigure}[t]{0.24\linewidth}
			\centering
			\includegraphics[width=1\columnwidth]{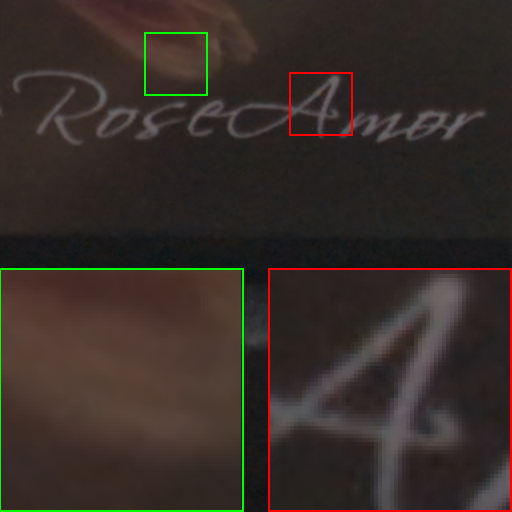}
			\caption*{DnCNN \cite{DnCNN} \\ 38.19/0.9051}
		\end{subfigure}
		\begin{subfigure}[t]{0.24\linewidth}
			\centering
			\includegraphics[width=1\columnwidth]{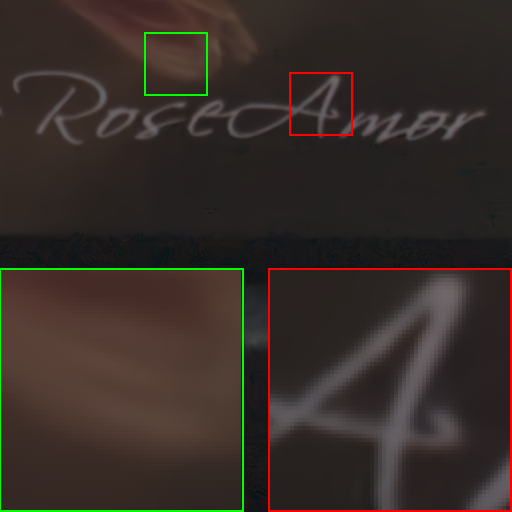}
			\caption*{FFDNet \cite{FFDNet} \\ 38.32/0.9083}
		\end{subfigure}
		\begin{subfigure}[t]{0.24\linewidth}
			\centering
			\includegraphics[width=1\columnwidth]{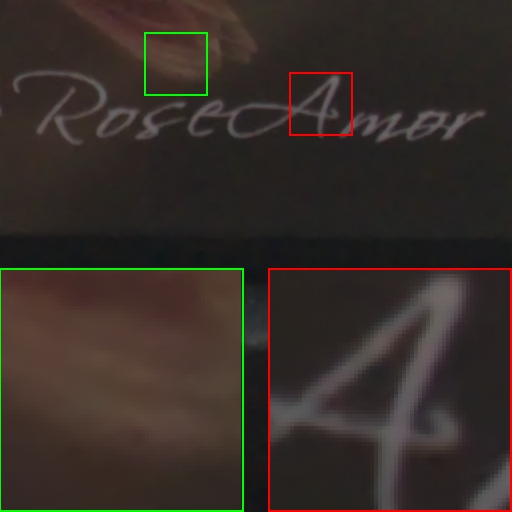}
			\caption*{CBDNet \cite{CBDNet} \\ 38.74/0.9132}
		\end{subfigure}
		\\
		\begin{subfigure}[t]{0.24\linewidth}
			\centering
			\includegraphics[width=1\columnwidth]{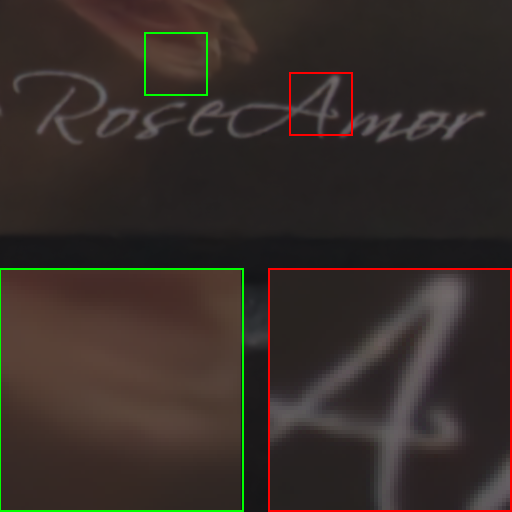}
			\caption*{RIDNet \cite{RIDNet} \\ 39.55/0.9228}
		\end{subfigure}
		\begin{subfigure}[t]{0.24\linewidth}
			\centering
			\includegraphics[width=1\columnwidth]{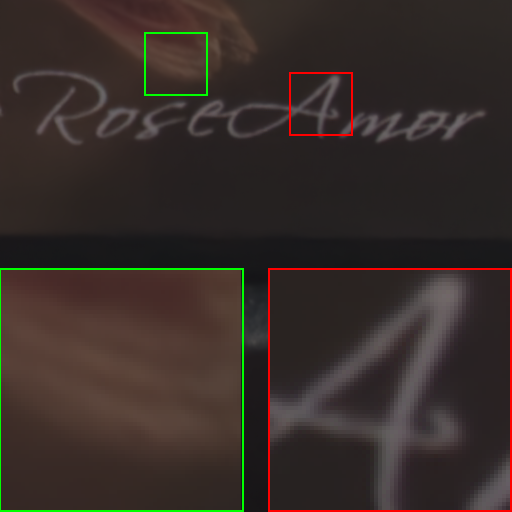}
			\caption*{VDN \cite{VDN} \\ 39.87/0.9277}
		\end{subfigure}
		\begin{subfigure}[t]{0.24\linewidth}
			\centering
			\includegraphics[width=1\columnwidth]{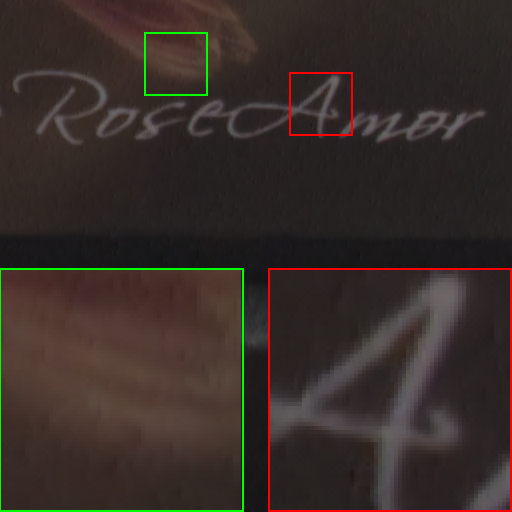}
			\caption*{AINDNet \cite{AINDNet} \\ 39.00/0.9148}
		\end{subfigure}
		\begin{subfigure}[t]{0.24\linewidth}
			\centering
			\includegraphics[width=1\columnwidth]{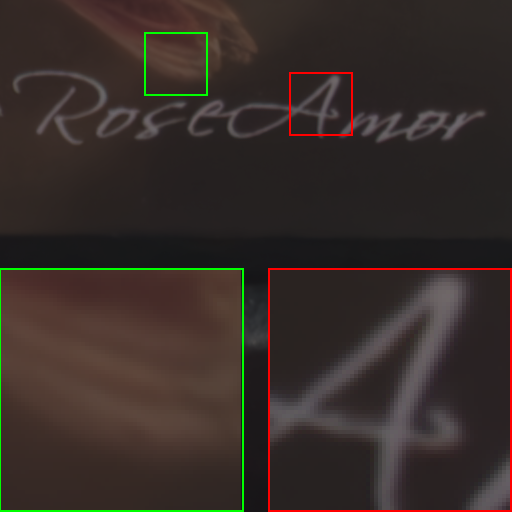}
			\caption*{VDID+ (Ours) \\ 39.98/0.9285}
		\end{subfigure}			
	\end{center}
	\vspace{-0.5cm}
	\caption{Visualized examples from DND benchmark \cite{DND} with PSNR/SSIM.}
	\label{fig:001}
\end{figure}

\begin{figure*}
	\begin{center}		
		\captionsetup{justification=centering}
		\begin{subfigure}[t]{0.16\linewidth}
			\centering
			\includegraphics[width=1\columnwidth]{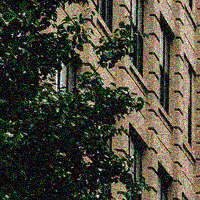}
			\caption{Noisy image $\sigma_N=30$ \\ 19.72 dB}
			\label{fig:latent1a}
		\end{subfigure}
		\begin{subfigure}[t]{0.16\linewidth}
			\centering
			\includegraphics[width=1\columnwidth]{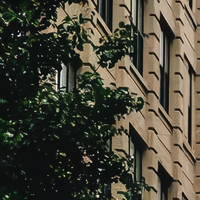}
			\caption{Restoration of (a) \\ 27.20 dB}
			\label{fig:latent1b}
		\end{subfigure}
		\begin{subfigure}[t]{0.16\linewidth}
			\centering
			\includegraphics[width=1\columnwidth]{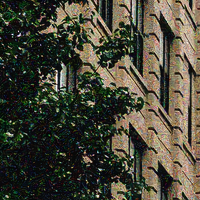}
			\caption{Restoration with \\ $\c$ from $\sigma_N=10$ \\ 23.95 dB}
			\label{fig:latent1c}
		\end{subfigure}
		\begin{subfigure}[t]{0.16\linewidth}
			\centering
			\includegraphics[width=1\columnwidth]{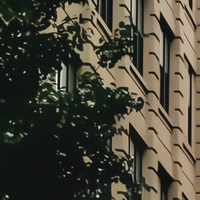}
			\caption{Restoration with $\c$ from $\sigma_N=50$ \\ 24.93 dB}
			\label{fig:latent1d}
		\end{subfigure}
		\begin{subfigure}[t]{0.16\linewidth}
			\centering
			\includegraphics[width=1\columnwidth]{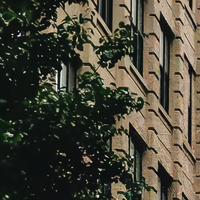}
			\caption{Restoration with \\ $\c$ from flipped input \\ 26.74 dB}
			\label{fig:latent1e}
		\end{subfigure}
\end{center}
	\vspace{-0.5cm}
	\caption{Denoising results depending on $\c$. We feed a noisy image (a) to the denoising network while changing the input to the encoder that extracts $\c$. (b) Result of VDID with the same input to the denoiser and encoder. (c) Result when the encoder is given the input with $\sigma_N=10$ so that
		$\c$ is the latent variable correponding to $\sigma_N=10$. (d) $\c$ correponds to $\sigma_N=50$. (e) The encoder is given the flipped patch with $\sigma_N=30$. Also see the visualized $\c$ in Fig.~\ref{fig:cond1}, and text (Section \ref{sec:analysis}) for details. 
}
	\label{fig:latent1}
\end{figure*}

\begin{figure}
		\begin{center}		
			\captionsetup{justification=centering}
			\begin{subfigure}[t]{0.24\linewidth}
				\centering
				\includegraphics[width=1\columnwidth]{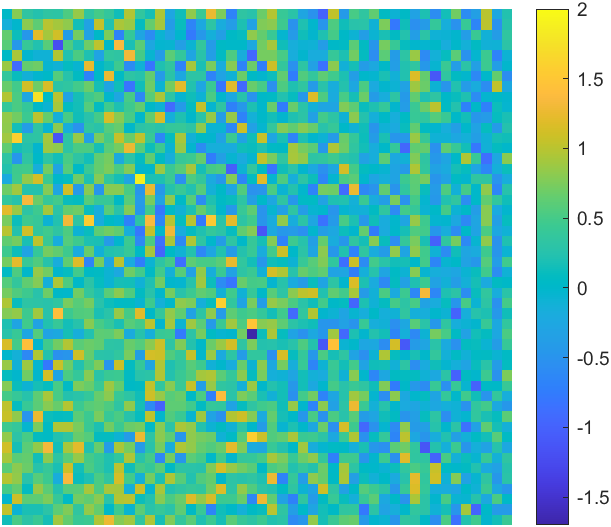}
				\caption{Encoder input with $\sigma_N=30$}
				\label{fig:cond1a}
			\end{subfigure}
			\begin{subfigure}[t]{0.24\linewidth}
				\centering
				\includegraphics[width=1\columnwidth]{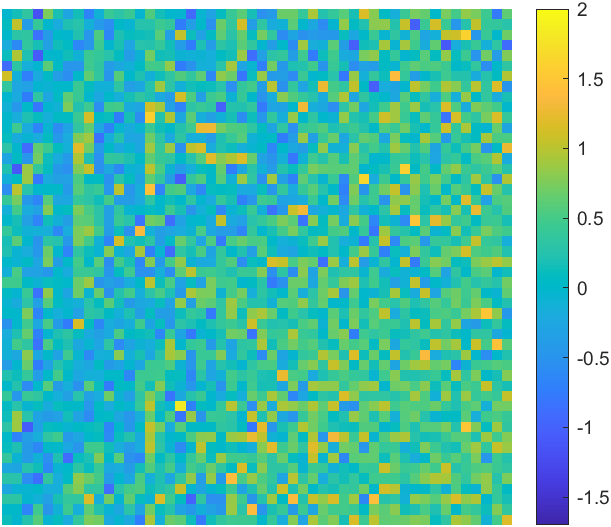}
				\caption{Flipped input, $\sigma_N=30$}
				\label{fig:cond1b}
			\end{subfigure}
			\begin{subfigure}[t]{0.24\linewidth}
				\centering
				\includegraphics[width=1\columnwidth]{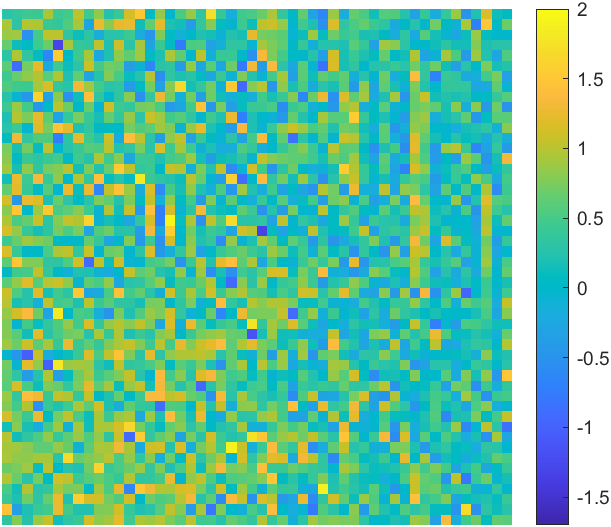}
				\caption{Input with $\sigma_N=10$}
				\label{fig:cond1c}
			\end{subfigure}
			\begin{subfigure}[t]{0.24\linewidth}
				\centering
				\includegraphics[width=1\columnwidth]{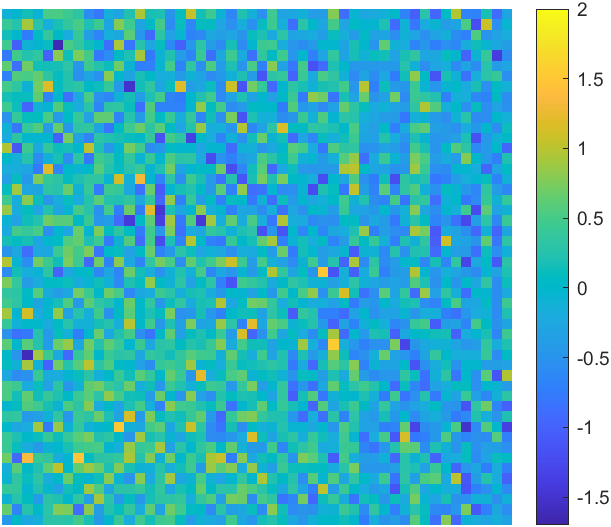}
				\caption{Input with $\sigma_N=50$}
				\label{fig:cond1d}
			\end{subfigure}
			\\
			\begin{subfigure}[t]{0.24\linewidth}
				\centering
				\includegraphics[width=1\columnwidth]{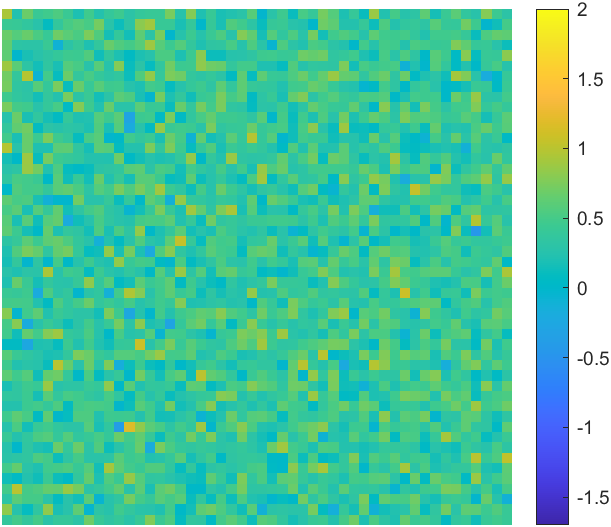}
				\caption{Flat input, $\sigma_N=10$}
				\label{fig:cond1e}
			\end{subfigure}
			\begin{subfigure}[t]{0.24\linewidth}
				\centering
				\includegraphics[width=1\columnwidth]{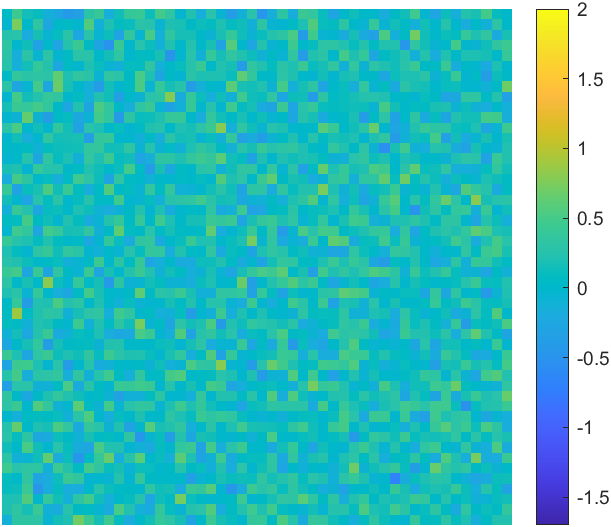}
				\caption{Flat input, $\sigma_N=30$}
				\label{fig:cond1f}
			\end{subfigure}
			\begin{subfigure}[t]{0.24\linewidth}
				\centering
				\includegraphics[width=1\columnwidth]{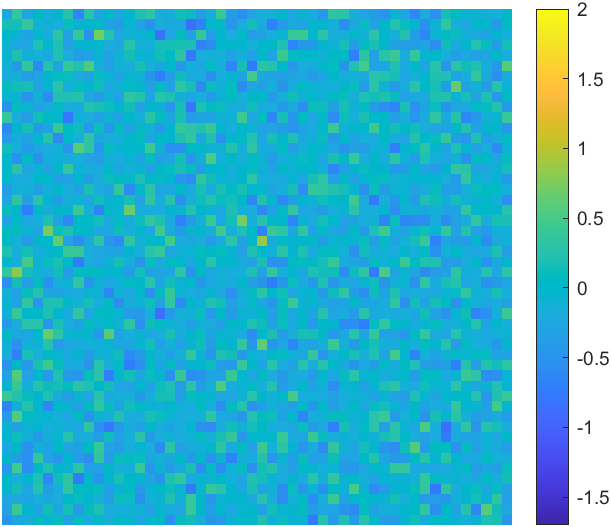}
				\caption{Flat input, $\sigma_N=50$}
				\label{fig:cond1g}
			\end{subfigure}		
		\end{center}
	\vspace{-0.5cm}
	\caption{Visualization of $\c$ for the image in \figurename{~\ref{fig:latent1}} and for a flat image with varying $\sigma_N$. (a) $\c$ from the encoder, when its input is Fig.~\ref{fig:latent1a}, (b) when the input is the flip of Fig.~\ref{fig:latent1a}, (c) when the input is the same but $\sigma_N=10$, and (d) $\sigma_N=50$. (e) $\c$ from the encoder given a flat patch with $\sigma_N=10$, (f) $\sigma_N=30$, and (g) $\sigma_N=50$. See text (Section \ref{sec:analysis}) for details.}
	\label{fig:cond1}
\end{figure}

\section{Analysis}
\label{sec:analysis}

To investigate the role of our latent variable $\c$, we conduct several experiments exhibited in \figurename{~\ref{fig:latent1} and \figurename{~\ref{fig:cond1}}.

\paragraph{Latent Variable Manipulation}
In our VDID, the denoiser and encoder should be given the same noisy image so that the encoder extracts appropriate $\c$ that bears information needed for the denoiser. That is, when the same image with $\sigma_N=30$ (\figurename{~\ref{fig:latent1a}}) is fed to the denoiser and encoder, the VDID performs the best as in \figurename{~\ref{fig:latent1b}. If we feed the same input ($\sigma_N=30$) to the denoiser while feeding input with $\sigma_N=10$ to the encoder, then the encoder will extract $\c$ corresponding to $\sigma_N=10$, and the performance is degraded as in 
\figurename{~\ref{fig:latent1c}. Similarly, feeding the image with $\sigma_N=50$ also lowers the PSNR (\figurename{~\ref{fig:latent1d}). To see that the performance is also affected by the change of image contents, we feed a flipped image of \figurename{~\ref{fig:latent1a} to the encoder so that the $\c$ does not match with denoiser input.
Although the noise level is the same, the PSNR is lowered as in \figurename{~\ref{fig:latent1e}, showing that $\c$ delivers image information as well as noise information. This is better illustrated in \figurename{~\ref{fig:cond1}} showing the visualized $\c$, which is also explained in the next paragraph. 

\vspace{-0.3cm}
\paragraph{Latent Variable Visualization}
To further investigate the latent space, we visualize the channel-wise average of $\c$ by feeding various inputs, as in \figurename{~\ref{fig:cond1}}. The first row shows $\c$ extracted from the encoder, when the input image is \figurename{~\ref{fig:latent1}} with $\sigma_N=30$, flipped image with $\sigma_N=30$, and the images with $\sigma_N=10$ and $50$. The second row shows $\c$ for the flat image with $\sigma_N=10$, $30$, and $50$ for the reference. By comparing the first and second rows, it can be seen that $\c$ bears image contents. Specifically, the flat image generates almost flat $\c$ with the changing magnitudes according to $\sigma_N$, validating that $\c$ delivers noise information. When the input has both texture and flat areas (the first row), then $\c$ shows the magnitudes changing with respect to contents and also noise variance. Comparing \figurename{~\ref{fig:cond1a}} and the second-row images, it is interesting to see that texture area (tree) pretends to have lower $\sigma_N$ (yellow as \figurename{~\ref{fig:cond1e}}), and flat areas (building) pretends to be higher as \figurename{~\ref{fig:cond1g}}. This means that the flat area is more strongly filtered than the textured. 

The visualizations also validate why we obtain the result of \figurename{~\ref{fig:latent1e}}, when we feed a flipped image to the encoder. Precisely, the tree area is over-smoothed and the building area is less filtered because $\c$ is conversely delivered. Likewise, providing noise-only information (flat $\c$) also lowers the performance as in Figs.~{~\ref{fig:latent1}}(c) and (d), which also imply that our learned $\c$ bears more information than the noise variance, for the successful denoising.

\vspace{-0.3cm}
\paragraph{Summary of Latent Variable Analysis}
In summary, we note that using only noise variance information is not enough for the successful denoising, which was a common approach in previous works \cite{DnCNN, FFDNet, ATDNet, DUBD}. Instead, the optimal method would be to use both noise and content information.
In our method, the encoder learns such information and delivers it to the denoiser as the latent variable $\c$, which eventually boosts the performance.

\vspace{-0.3cm}
\paragraph{Ablation Study}
As we propose to use additional loss terms, we provide ablation results on loss terms in \tablename{~\ref{tab:ablation}}. With only denoising loss term $\mathcal{L}_{denoise}$, which corresponds to na\"{i}ve blind denoising, the performance is inferior to other methods.
Without the adversarial loss $\mathcal{L}_{adv}$ for approximating the likelihood, it shows a slightly inferior result compared to using total loss because using only the pixel-wise distance strictly assumes a probability distribution family. Interestingly, during our research, we found that without the adversarial loss, the encoder mainly encodes the content information such as color, while the noise level information is rarely embedded. By using all loss terms, the proposed method guarantees the performance that surpasses others.

\begin{table}
	\caption{Ablation study on CBSD68 with $\sigma_N=30$.}
	\label{tab:ablation}
	\centering
	\begin{tabular}{l|c}
		\hline
		\rule[-1ex]{0pt}{3.5ex}
		Loss term     & PSNR \\
		\hline
		\rule[-1ex]{0pt}{3.5ex}
		$\mathcal{L}_{denoise}$ & 30.47 \\
		\rule[-1ex]{0pt}{3.5ex}
		$\mathcal{L}_{denoise} + \beta D_{KL} + \mathcal{L}_{recon}$ w/o $\mathcal{L}_{adv}$ & 30.56 \\
		\rule[-1ex]{0pt}{3.5ex}
		$\mathcal{L}_{denoise} + \beta D_{KL} + \mathcal{L}_{recon}$ w/ $\mathcal{L}_{adv}$ & 30.64 \\
		\hline
	\end{tabular}
\end{table}

\section{Conclusion}
In this paper, we have presented a novel variational approach for image denoising. Concretely, we reformulated the log-posterior of the denoising problem and introduced a \emph{variational lower bound} to approximate the original MAP objective. With our \emph{variational lower bound}, the original problem can be divided into separate sub-problems, which eventually relaxes the given problem. Moreover, our \emph{variational lower bound} incorporates both denoising objective and noisy image generative models. Hence, we can further exploit the embedded information of complicated noisy image manifold. We have also presented three parameterized CNNs for the inference problem and shown that our method achieves state-of-the-art performance in removing both Gaussian and real-world noises while requiring fewer parameters. The code is publicly available at \url{https://github.com/JWSoh/VDIR}.

\clearpage
\onecolumn
\section*{Appendix}

\section*{A. Derivation and Proof of the Variational Lower Bound}
	
Given a noisy image $\y$ and its underlying clean image $\x$, the joint distribution can be reformulated as follows.
We introduce a tractable distribution $q(\c|\y)$ and a latent variable $\c$.

\begin{align}\nonumber
\log p(\x, \y) =& \int_{c} q(\c|\y) \log p(\x, \y) d\c \\\nonumber
=& \int_{c} q(\c|\y) \log \frac{p(\x|\y,\c) p(\y, \c)}{p(\c|\x,\y)} d\c\\\nonumber
=& \int_{c} q(\c|\y) \log \frac{p(\x|\y,\c) p(\y|\c) p(\c)}{p(\c|\x,\y)} d\c\\\nonumber
=& \int_{c} q(\c|\y) [\log p(\x|\y,\c) +\log \frac{q(\c|\y)}{p(\c|\x,\y)} + \log \frac{p(\c)}{q(\c|\y)} + \log p(\y|\c)]d\c\\\nonumber
=& \mathbb{E}_{\c \sim q(\c|\y)} [\log p(\x|\y,\c)] + D_{KL}(q(\c|\y)||p(\c|\x,\y)) \\\nonumber
&\qquad \qquad - D_{KL}(q(\c|\y)||p(\c)) + \mathbb{E}_{\c \sim q(\c|\y)} [\log p(\y|\c)].
\end{align}
The \emph{variational lower bound} is defined as
\begin{equation}
\mathcal{L} = \mathbb{E}_{\c \sim q(\c|\y)} [\log p(\x|\y,\c)] - D_{KL}(q(\c|\y)||p(\c)) + \mathbb{E}_{\c \sim q(\c|\y)} [\log p(\y|\c)],
\end{equation}
then,
\begin{equation}
\log p(\x, \y) = \mathcal{L} + D_{KL}(q(\c|\y)||p(\c|\x,\y)).
\end{equation}
Since the KL divergence is non-negative, the following inequality holds.
\begin{equation}
\log p(\x,\y) \geq \mathcal{L}.
\end{equation}

\section*{B. The Auto-Encoder \& Discriminator Architecture}
\tablename{s~\ref{tab:encoder}} and \ref{tab:decoder} show the network architectures of the encoder and decoder, respectively. The number of parameters of the encoder is $154$ K and the decoder is $114$ K. Also, \tablename{~\ref{tab:dis}} describes the discriminator architecture.  We adopt patch discriminator \cite{Pix2Pix}, and spectral normalization \cite{SNGAN} is used for all convolution layers in the discriminator.
The notations are as follows.
\begin{itemize}
	\item $H, W, C$: Height, width, and the number of channels of the input image.
	\item Conv2d(K, S): 2d convolution with kernel size K and stride S.
	\item MaxPool(K, S): Max pooling operation with kernel size K and strides S.
	\item NN Upsampling: Nearest neighbor upsampling.
\end{itemize}

\begin{table}
	\caption{The network architecture of the encoder.}
	\label{tab:encoder}
	\centering
	\begin{tabular}{l l l l}
		\Xhline{4\arrayrulewidth}
		\rule[-1ex]{0pt}{3.5ex}
		Module & Layers & Input size & Output size \\
		\hline
		\rule[-1ex]{0pt}{3.5ex}
		\emph{Block 1} & Conv2d(3, 1) & $H\times W\times C$ & $H\times W\times 64$ \\
		\rule[-1ex]{0pt}{3.5ex}
		& MaxPool(2, 2), ReLU & $H\times W\times 64$ & $\frac{H}{2}\times \frac{W}{2} \times 64$ \\
		\hline
		\rule[-1ex]{0pt}{3.5ex}
		\emph{Block 2} & Conv2d(3, 1), ReLU &  $\frac{H}{2}\times \frac{W}{2} \times 64$ &  $\frac{H}{2}\times \frac{W}{2} \times 64$ \\
		\rule[-1ex]{0pt}{3.5ex}
		& Conv2d(3, 1) &  $\frac{H}{2}\times \frac{W}{2} \times 64$ &  $\frac{H}{2}\times \frac{W}{2} \times 64$ \\
		\rule[-1ex]{0pt}{3.5ex}
		& MaxPool(2, 2), ReLU &  $\frac{H}{2}\times \frac{W}{2} \times 64$ &  $\frac{H}{4}\times \frac{W}{4} \times 64$ \\
		\hline
		\rule[-1ex]{0pt}{3.5ex}
		\emph{Block 3} & Conv2d(3, 1), ReLU &  $\frac{H}{4}\times \frac{W}{4} \times 64$ &  $\frac{H}{4}\times \frac{W}{4} \times 64$ \\
		\rule[-1ex]{0pt}{3.5ex}
		(a) & Conv2d(3, 1) &  $\frac{H}{4}\times \frac{W}{4} \times 64$ &  $\frac{H}{4}\times \frac{W}{4} \times 64$ \\
		\hline
		\rule[-1ex]{0pt}{3.5ex}
		\emph{$\mu$}, input (a) & Conv2d(3, 1) &  $\frac{H}{4}\times \frac{W}{4} \times 64$ &  $\frac{H}{4}\times \frac{W}{4} \times 4$ \\
		\rule[-1ex]{0pt}{3.5ex}
		\emph{$\log \sigma^2$}, input (a)& Conv2d(3, 1) &  $\frac{H}{4}\times \frac{W}{4} \times 64$ &  $\frac{H}{4}\times \frac{W}{4} \times 4$ \\

		\Xhline{4\arrayrulewidth}
	\end{tabular}
\end{table}

\begin{table}
	\caption{The network architecture of the decoder.}
	\label{tab:decoder}
	\centering
	\begin{tabular}{l l l l}
		\Xhline{4\arrayrulewidth}
		\rule[-1ex]{0pt}{3.5ex}
		Module & Layers & Input size & Output size \\
		\hline
		\rule[-1ex]{0pt}{3.5ex}
		\emph{Block 1} & Conv2d(3, 1) &$\frac{H}{4}\times \frac{W}{4}\times 4$ &$\frac{H}{4}\times \frac{W}{4}\times 64$ \\
		\rule[-1ex]{0pt}{3.5ex}
		& NN Upsampling ($\times2$), ReLU& $\frac{H}{4}\times \frac{W}{4}\times 64$ &$\frac{H}{2}\times \frac{W}{2}\times 64$ \\
		\hline
		\rule[-1ex]{0pt}{3.5ex}
		\emph{Block 2} & Conv2d(3, 1), ReLU &  $\frac{H}{2}\times \frac{W}{2} \times 64$ &  $\frac{H}{2}\times \frac{W}{2} \times 64$ \\
		\rule[-1ex]{0pt}{3.5ex}
		& Conv2d(3, 1) &$\frac{H}{2}\times \frac{W}{2}\times 4$ &$\frac{H}{2}\times \frac{W}{2}\times 64$ \\
		\rule[-1ex]{0pt}{3.5ex}
		& NN Upsampling ($\times2$), ReLU& $\frac{H}{2}\times \frac{W}{2}\times 64$ &$H \times W\times 64$ \\
		\hline
		\rule[-1ex]{0pt}{3.5ex}
		\emph{Block 3} & Conv2d(3, 1), ReLU &  $ H \times W \times 64$ &  $H\times W \times 64$ \\
		\hline
		\rule[-1ex]{0pt}{3.5ex}
		\emph{Output} & Conv2d(3, 1) & $ H \times W \times 64$ &  $H\times W \times C$ \\	
		\Xhline{4\arrayrulewidth}
	\end{tabular}
\end{table}

\begin{table}
	\caption{The network architecture of the discriminator.}
	\label{tab:dis}
	\centering
	\begin{tabular}{l l l l}
		\Xhline{4\arrayrulewidth}
		\rule[-1ex]{0pt}{3.5ex}
		Module & Layers & Input size & Output size \\
		\hline
		\rule[-1ex]{0pt}{3.5ex}
		\emph{Conv 1\_1} & Conv2d(3, 1), leakyReLU & $H\times W\times C$ & $H\times W\times 64$ \\
		\rule[-1ex]{0pt}{3.5ex}
		\emph{Conv 1\_2} & Conv2d(3, 2), leakyReLU & $H\times W\times 64$ &  $\frac{H}{2}\times \frac{W}{2}\times 64$ \\
		\hline
		\rule[-1ex]{0pt}{3.5ex}
		\emph{Conv 2\_1} & Conv2d(3, 1), leakyReLU & $\frac{H}{2}\times \frac{W}{2}\times 64$ & $\frac{H}{2}\times \frac{W}{2}\times 128$ \\
		\rule[-1ex]{0pt}{3.5ex}
		\emph{Conv 2\_2} & Conv2d(3, 2), leakyReLU & $\frac{H}{2}\times \frac{W}{2}\times 128$ & $\frac{H}{4}\times \frac{W}{4}\times 128$ \\
		\hline
		\rule[-1ex]{0pt}{3.5ex}
		\emph{Conv 3\_1} & Conv2d(3, 1), leakyReLU & $\frac{H}{4}\times \frac{W}{4}\times 128$ & $\frac{H}{4}\times \frac{W}{4}\times 256$ \\
		\rule[-1ex]{0pt}{3.5ex}
		\emph{Conv 3\_2} & Conv2d(3, 2), leakyReLU & $\frac{H}{4}\times \frac{W}{4}\times 256$ & $\frac{H}{8}\times \frac{W}{8}\times 256$ \\
		\hline
		\rule[-1ex]{0pt}{3.5ex}
		\emph{Conv 4\_1} & Conv2d(3, 1), leakyReLU & $\frac{H}{8}\times \frac{W}{8}\times 256$ & $\frac{H}{8}\times \frac{W}{8}\times 512$ \\
		\rule[-1ex]{0pt}{3.5ex}
		\emph{Conv 4\_2} & Conv2d(3, 2), leakyReLU & $\frac{H}{8}\times \frac{W}{8}\times 512$ & $\frac{H}{16}\times \frac{W}{16}\times 512$ \\
		\hline
		\rule[-1ex]{0pt}{3.5ex}
		\emph{Conv 5\_1} & Conv2d(3, 1), leakyReLU & $\frac{H}{16}\times \frac{W}{16}\times 512$ & $\frac{H}{16}\times \frac{W}{16}\times 512$ \\
		\rule[-1ex]{0pt}{3.5ex}
		\emph{Conv 5\_2} & Conv2d(3, 2), leakyReLU & $\frac{H}{16}\times \frac{W}{16}\times 512$ & $\frac{H}{32}\times \frac{W}{32}\times 512$ \\
		
		\hline
		\rule[-1ex]{0pt}{3.5ex}
		\emph{logits} & Conv2d(3, 1) &  $\frac{H}{32}\times \frac{W}{32} \times 512$ &  $\frac{H}{32}\times \frac{W}{32} \times 1$ \\
		
		\Xhline{4\arrayrulewidth}
	\end{tabular}
\end{table}

\newpage
\section*{C. More Visualized Results}
We present additional visualized results in \figurename{s~\ref{fig:001_1},~\ref{fig:002},~\ref{fig:003},} and \ref{fig:004}.

\begin{figure}
	\begin{center}
		\begin{subfigure}[t]{0.19\linewidth}
			\centering
			\includegraphics[width=1\columnwidth]{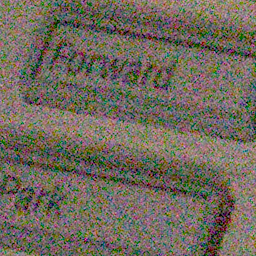}
			\caption*{Noisy}
		\end{subfigure}
		\begin{subfigure}[t]{0.19\linewidth}
			\centering
			\includegraphics[width=1\columnwidth]{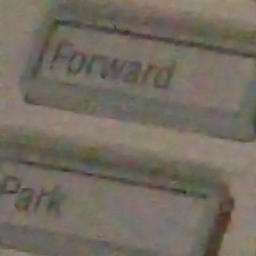}
			\caption*{CBDNet \cite{CBDNet} / 4.4 M}
		\end{subfigure}
		\begin{subfigure}[t]{0.19\linewidth}
			\centering
			\includegraphics[width=1\columnwidth]{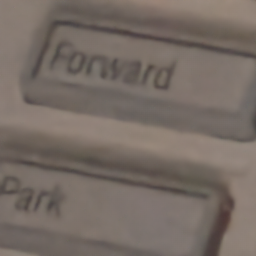}
			\caption*{RIDNet \cite{RIDNet} / 1.5 M}
		\end{subfigure}
		\begin{subfigure}[t]{0.19\linewidth}
			\centering
			\includegraphics[width=1\columnwidth]{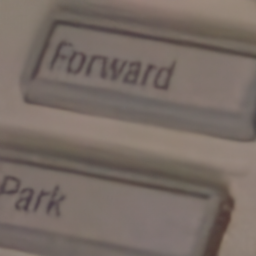}
			\caption*{VDN \cite{VDN} / 7.8 M}
		\end{subfigure}
		\begin{subfigure}[t]{0.19\linewidth}
			\centering
			\includegraphics[width=1\columnwidth]{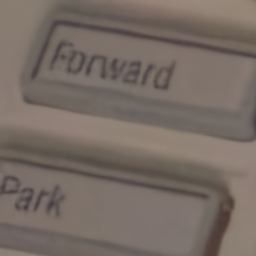}
			\caption*{VDID (Ours) / 2.2 M}
		\end{subfigure}
		
	\end{center}
	\caption{Visualized examples from SIDD validation set \cite{SIDD} with the number of parameters.}
	\label{fig:001_1}
\end{figure}

\begin{figure}
	\begin{center}
		\begin{subfigure}[t]{0.19\linewidth}
			\centering
			\includegraphics[width=1\columnwidth]{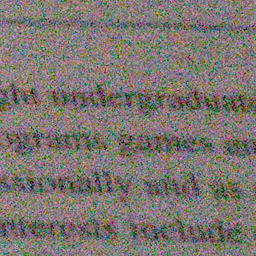}
			\caption*{Noisy}
		\end{subfigure}
		\begin{subfigure}[t]{0.19\linewidth}
			\centering
			\includegraphics[width=1\columnwidth]{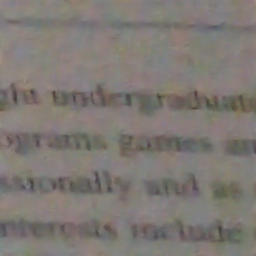}
			\caption*{CBDNet \cite{CBDNet} / 4.4 M}
		\end{subfigure}
		\begin{subfigure}[t]{0.19\linewidth}
			\centering
			\includegraphics[width=1\columnwidth]{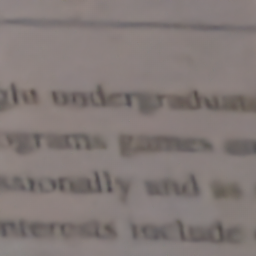}
			\caption*{RIDNet \cite{RIDNet} / 1.5 M}
		\end{subfigure}
		\begin{subfigure}[t]{0.19\linewidth}
			\centering
			\includegraphics[width=1\columnwidth]{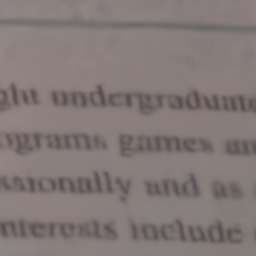}
			\caption*{VDN \cite{VDN} / 7.8 M}
		\end{subfigure}
		\begin{subfigure}[t]{0.19\linewidth}
			\centering
			\includegraphics[width=1\columnwidth]{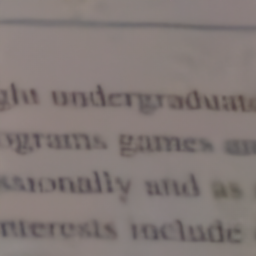}
			\caption*{VDID (Ours) / 2.2 M}
		\end{subfigure}
		
	\end{center}
	\caption{Visualized examples from SIDD validation set \cite{SIDD} with the number of parameters.}
	\label{fig:002}
\end{figure}
\begin{figure}
	
	\captionsetup{justification=centering}
	\begin{center}
		\begin{subfigure}[t]{0.216\linewidth}
			\centering
			\includegraphics[width=1\columnwidth]{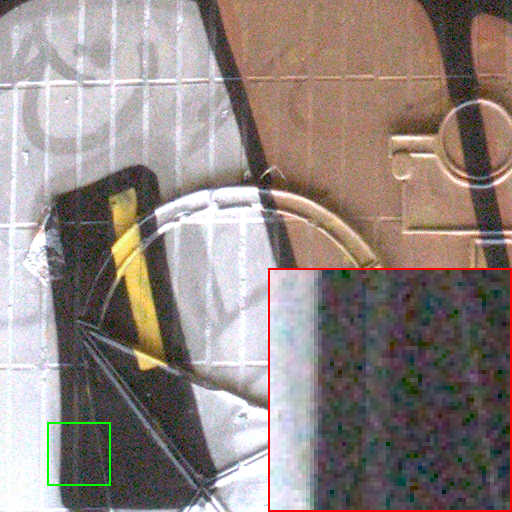}
			\caption*{Noisy \\ 26.90/0.7527}
		\end{subfigure}
		\begin{subfigure}[t]{0.216\linewidth}
			\centering
			\includegraphics[width=1\columnwidth]{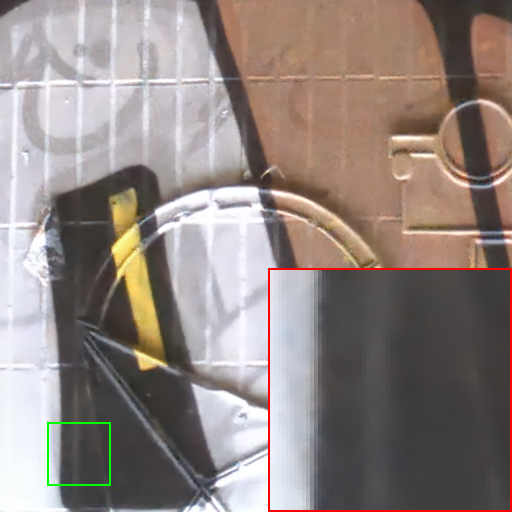}
			\caption*{DnCNN+ \cite{DnCNN} \\ 33.29/0.9271}
		\end{subfigure}
		\begin{subfigure}[t]{0.216\linewidth}
			\centering
			\includegraphics[width=1\columnwidth]{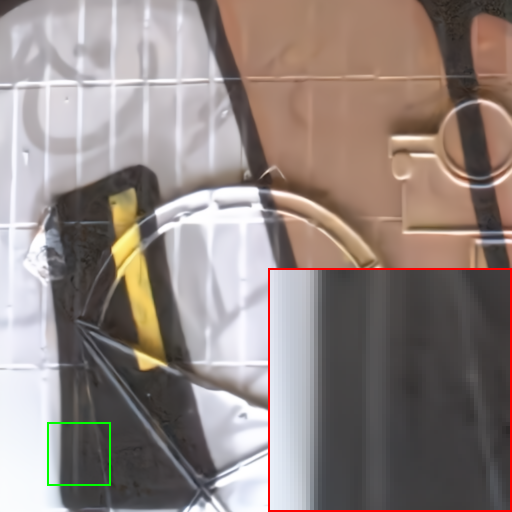}
			\caption*{FFDNet+ \cite{FFDNet} \\ 33.05/0.9231}
		\end{subfigure}
		\begin{subfigure}[t]{0.216\linewidth}
			\centering
			\includegraphics[width=1\columnwidth]{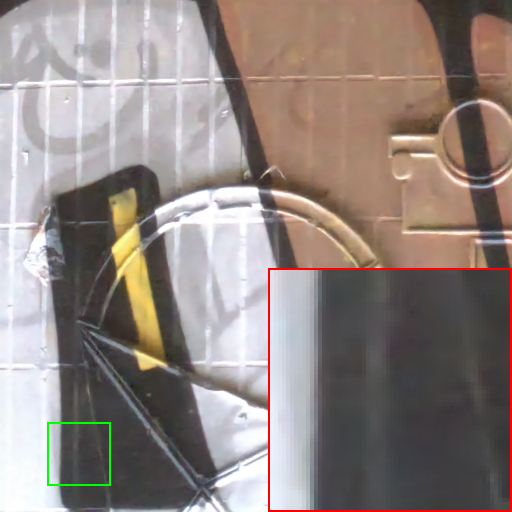}
			\caption*{CBDNet \cite{CBDNet} \\ 33.62/0.9295}
		\end{subfigure}
		\\
		\begin{subfigure}[t]{0.216\linewidth}
			\centering
			\includegraphics[width=1\columnwidth]{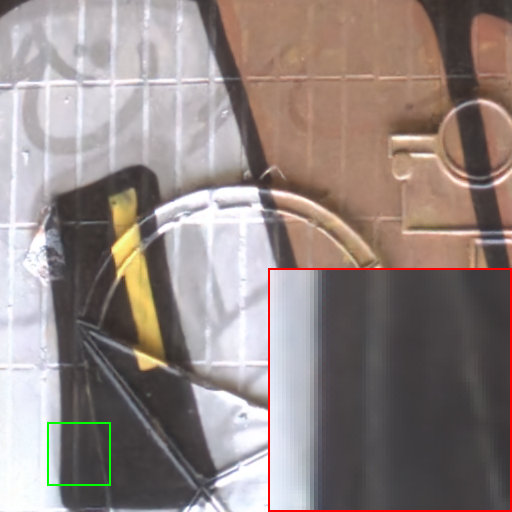}
			\caption*{RIDNet \cite{RIDNet} \\ 34.09/0.9382}
		\end{subfigure}
		\begin{subfigure}[t]{0.216\linewidth}
			\centering
			\includegraphics[width=1\columnwidth]{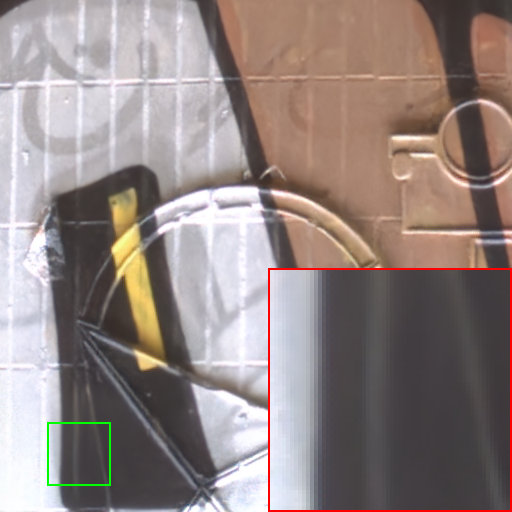}
			\caption*{VDN \cite{VDN} \\ 33.89/0.9376}
		\end{subfigure}
		\begin{subfigure}[t]{0.216\linewidth}
			\centering
			\includegraphics[width=1\columnwidth]{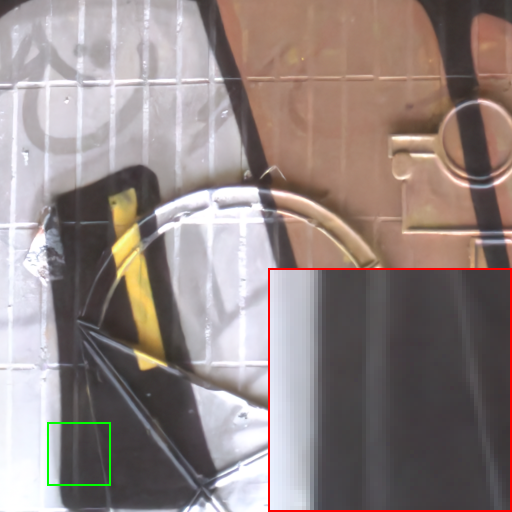}
			\caption*{AINDNet(S) \cite{AINDNet} \\ 34.28/0.9414}
		\end{subfigure}
		\begin{subfigure}[t]{0.216\linewidth}
			\centering
			\includegraphics[width=1\columnwidth]{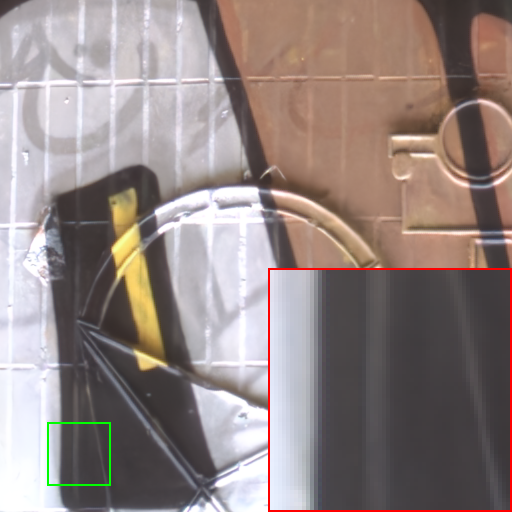}
			\caption*{VDID+ (Ours) \\ 34.63/0.9438}
		\end{subfigure}			
	\end{center}
	\caption{Visualized examples from DND benchmark \cite{DND} with PSNR/SSIM.}
	\label{fig:003}
\end{figure}

\begin{figure}	
	\captionsetup{justification=centering}
	\begin{center}
		\begin{subfigure}[t]{0.32\linewidth}
			\centering
			\includegraphics[width=1\columnwidth]{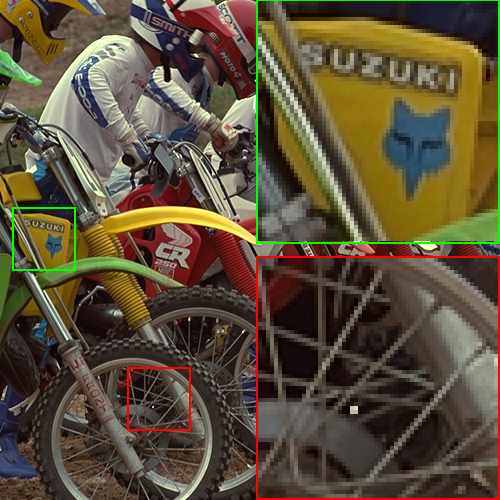}
			\caption*{GT \\ -}
		\end{subfigure}
		\begin{subfigure}[t]{0.32\linewidth}
			\centering
			\includegraphics[width=1\columnwidth]{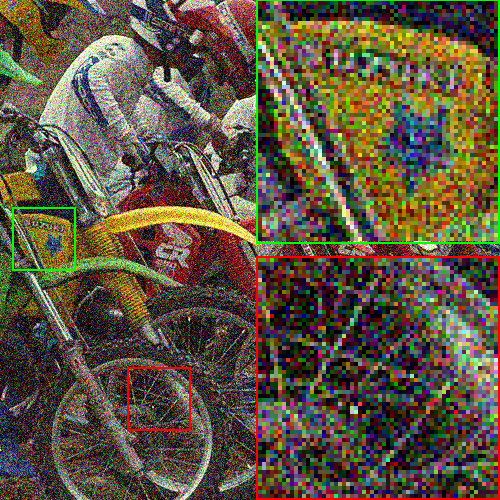}
			\caption*{Noisy \\ 15.11 dB}
		\end{subfigure}
		\begin{subfigure}[t]{0.32\linewidth}
			\centering
			\includegraphics[width=1\columnwidth]{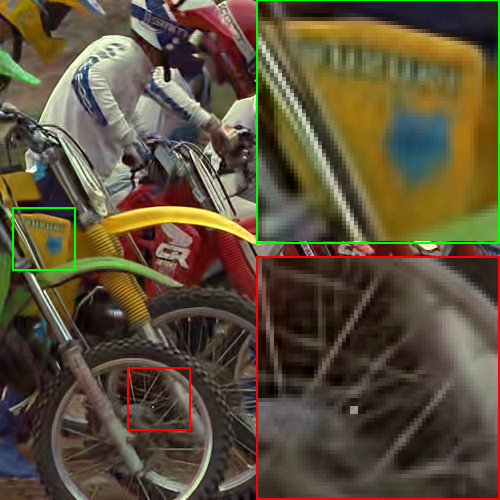}
			\caption*{CBM3D \cite{BM3D}\\ 25.50 dB}
		\end{subfigure}
		\begin{subfigure}[t]{0.32\linewidth}
			\centering
			\includegraphics[width=1\columnwidth]{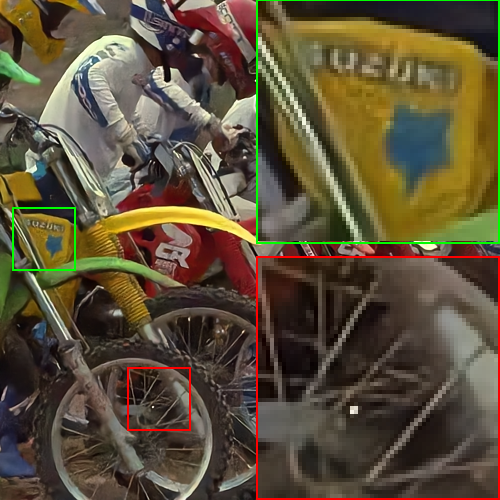}
			\caption*{DnCNN \cite{DnCNN} \\ 26.19 dB}
		\end{subfigure}
		\begin{subfigure}[t]{0.32\linewidth}
			\centering
			\includegraphics[width=1\columnwidth]{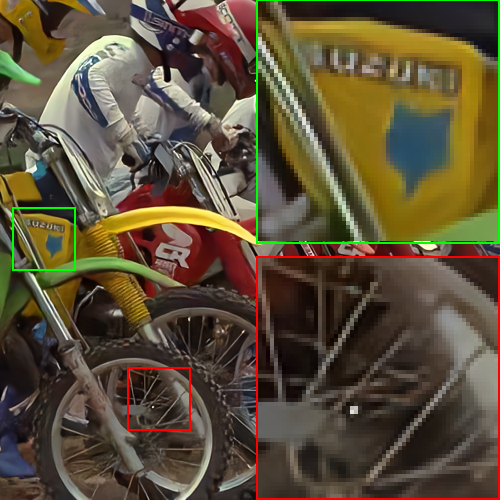}
			\caption*{FFDNet \cite{FFDNet} \\ 26.28 dB}
		\end{subfigure}
		\begin{subfigure}[t]{0.32\linewidth}
			\centering
			\includegraphics[width=1\columnwidth]{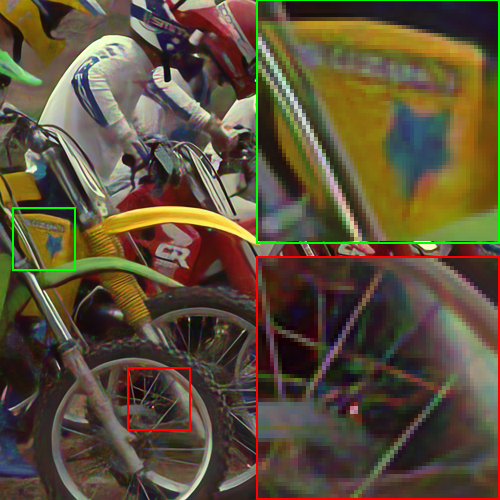}
			\caption*{RED \cite{RED} \\ 24.29 dB}
		\end{subfigure}
		\begin{subfigure}[t]{0.32\linewidth}
			\centering
			\includegraphics[width=1\columnwidth]{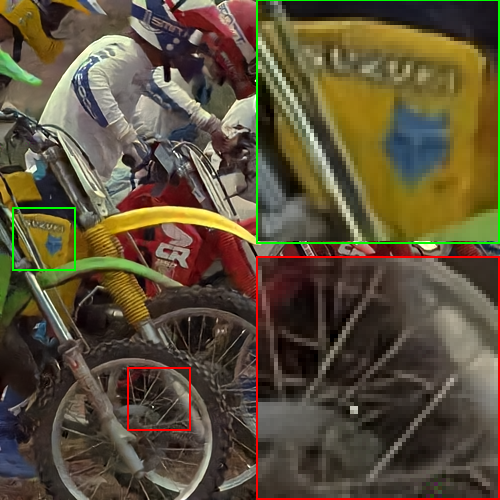}
			\caption*{UNLNet \cite{UNLNet} \\ 26.26 dB}
		\end{subfigure}
		\begin{subfigure}[t]{0.32\linewidth}
			\centering
			\includegraphics[width=1\columnwidth]{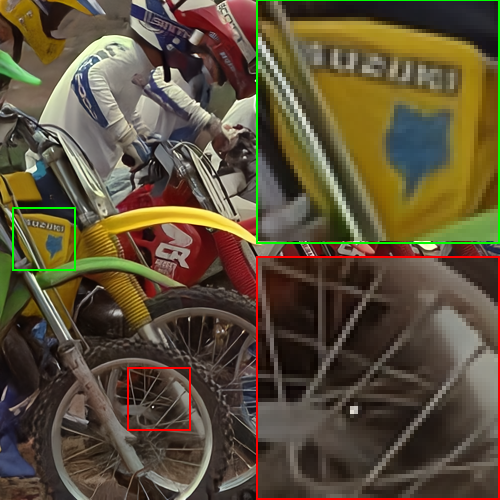}
			\caption*{VDN \cite{VDN} \\ 26.78 dB}
		\end{subfigure}
		\begin{subfigure}[t]{0.32\linewidth}
			\centering
			\includegraphics[width=1\columnwidth]{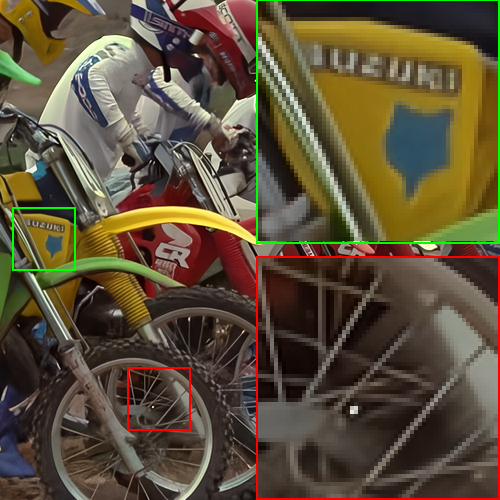}
			\caption*{VDID (Ours) \\ 26.89 dB}
		\end{subfigure}			
	\end{center}
	\caption{Visualized examples on Gaussian noise with $\sigma_N=50$ with PSNR.}
	\label{fig:004}
\end{figure}

\clearpage
\section*{D. More Analysis on Latent Space}

\paragraph{t-SNE Visualization}
For further analysis, we demonstrate t-SNE \cite{tSNE} visualization of $\c$ in \figurename{~\ref{fig:tSNE}} to investigate the latent space. In particular, $\c$ is average-pooled to generate global abstract of a patch as $AvgPool(\c) \in \mathbb{R}^4$. We first present results on Gaussian noise with DIV2K validation set \cite{DIV2K}, where $1,000$ patches are randomly sampled from DIV2K validation set, and Gaussian noises are added with noise levels $10$ to $55$ at $5$ intervals, in \figurename{~\ref{fig:tSNE_image}, ~\ref{fig:tSNE_noise}}.

As shown, the latent space well represents the noisy image manifold based on their contents and noise distribution. Interestingly, the latent embedding represents the content information and varies continuously as shown in \figurename{~\ref{fig:tSNE_image}}, in that similar color and luminance brought similar $\c$. The latent space also contains noise information as shown in \figurename{~\ref{fig:tSNE_noise}}. Specifically, the patches with similar noise levels are closely located and clustered. In conclusion, the latent space, which is suitable for denoising tasks, contains not only noise level information but also global content information.

Moreover, we also present t-SNE visualization of the real-noise denoising in \figurename{~\ref{fig:tSNE2_image}, ~\ref{fig:tSNE2_data}}. We extracted patches from three datasets: SIDD \cite{SIDD} validation set, DND \cite{DND}, and DIV2K validation set \cite{DIV2K} with synthetic noise following \cite{CBDNet}.

As shown, the latent embedding is highly correlated to the content information such as colors and intensities. It might be connected to the common knowledge that the noise from the real-world is signal-dependent. Interestingly, based on \figurename{~\ref{fig:tSNE2_data}}, the latent code captures different characteristics between the noise from SIDD\cite{SIDD} and DND\cite{DND}, despite we did not inject any supervision about the dataset. Concretely, they are separately clustered, and we might infer that there exists a domain gap between them.
In other words, our latent variable sees the difference in noise distribution of smartphone cameras and commercial cameras. Rather, the synthetic noise based on \cite{CBDNet} may better mimic the characteristics of the noise from commercial cameras compared to SIDD \cite{SIDD} based on our observation.

\begin{figure}
	\begin{center}
		\begin{subfigure}[t]{0.24\linewidth}
			\centering
			\includegraphics[width=1\columnwidth]{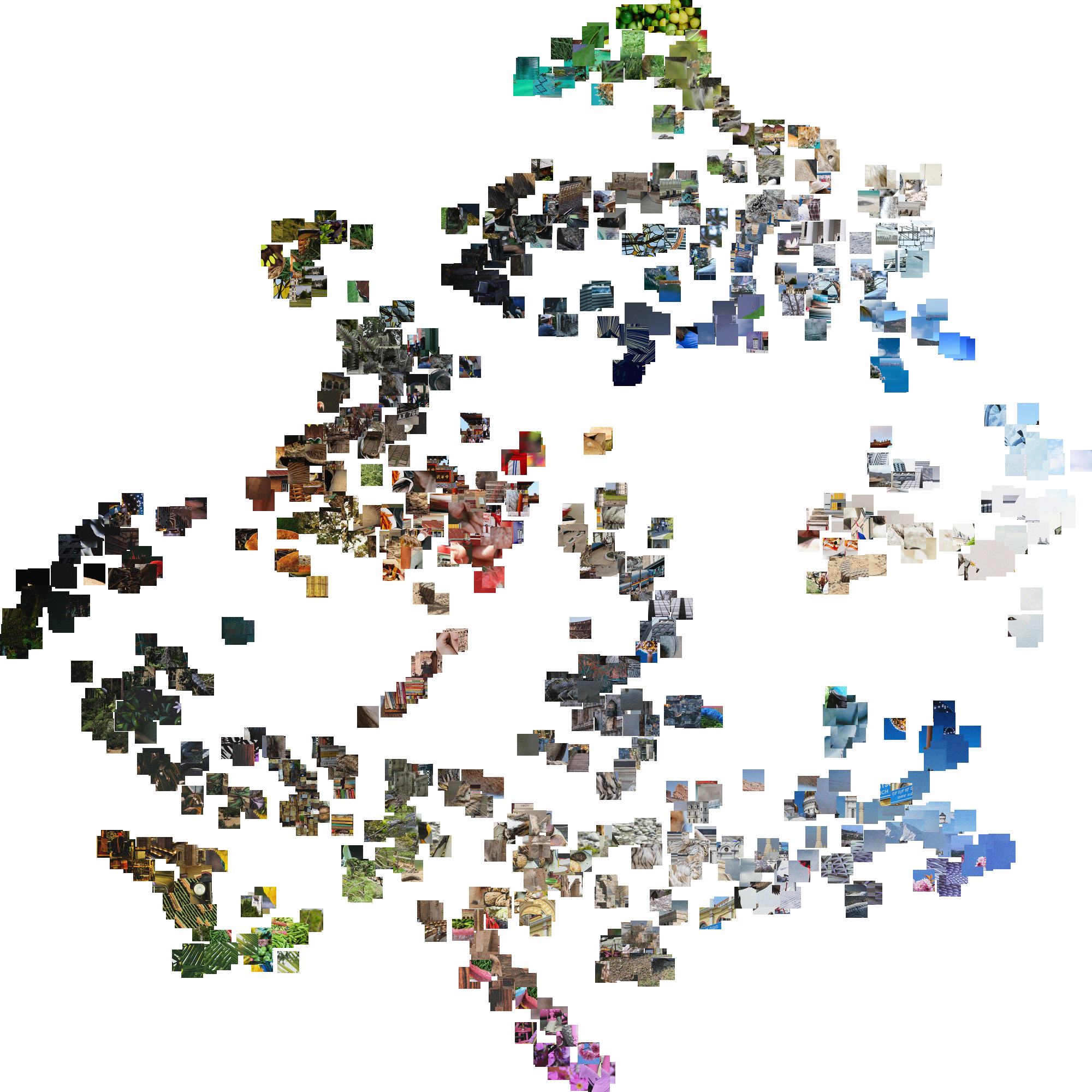}
			\caption{Patches (AWGN)}
			\label{fig:tSNE_image}
		\end{subfigure}
		\begin{subfigure}[t]{0.24\linewidth}
			\centering
			\includegraphics[width=1\columnwidth]{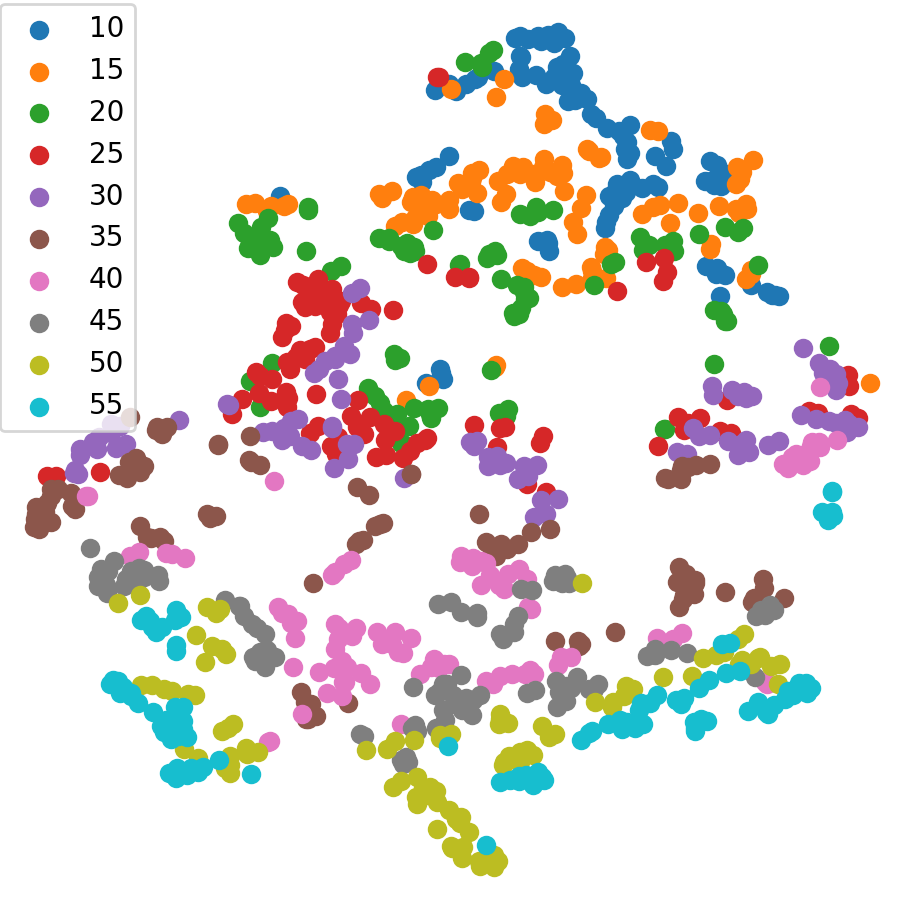}
			\caption{Noise levels (AWGN)}
			\label{fig:tSNE_noise}
		\end{subfigure}		
		\begin{subfigure}[t]{0.24\linewidth}
			\centering
			\includegraphics[width=1\columnwidth]{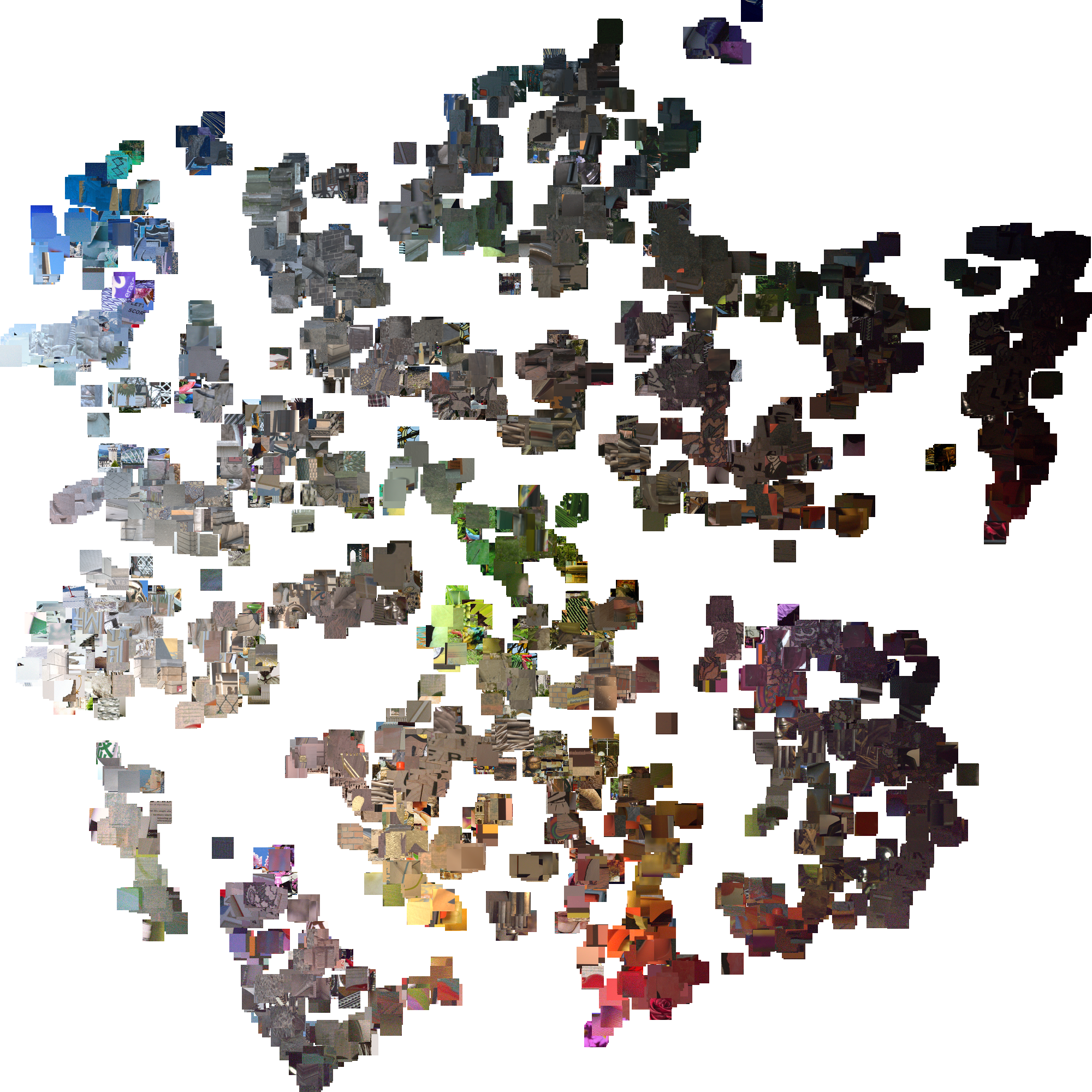}
			\caption{Patches (Real)}
			\label{fig:tSNE2_image}
		\end{subfigure}
		\begin{subfigure}[t]{0.24\linewidth}
			\centering
			\includegraphics[width=1\columnwidth]{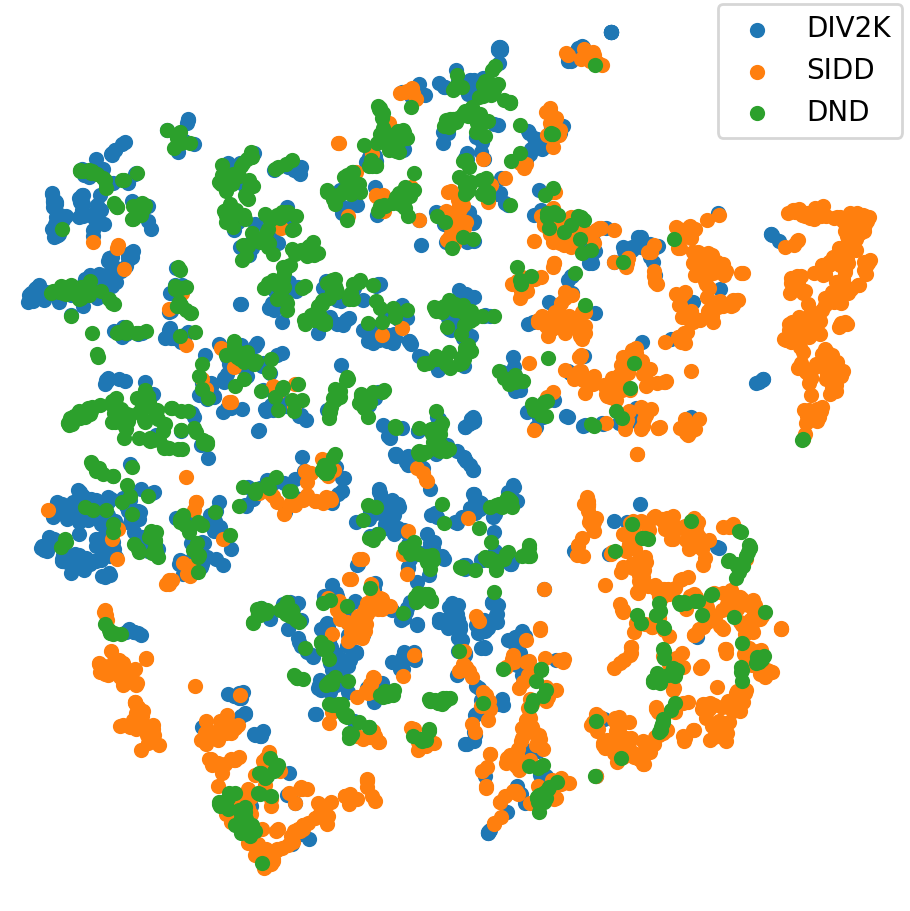}
			\caption{Datasets (Real)}
			\label{fig:tSNE2_data}
		\end{subfigure}
		
	\end{center}
	\caption{t-SNE visualizations on Gaussian (a-b) and real noise (c-d). (a) t-SNE visualization with patches. (b) Corresponding noise levels presented with different colors. (c) t-SNE visualization with patches. (d) Corresponding datasets.}
	\label{fig:tSNE}
\end{figure}

\end{document}